\documentclass{article}
\usepackage{subcaption}

\usepackage{arxiv}

\usepackage[utf8]{inputenc} % allow utf-8 input
\usepackage[T1]{fontenc}    % use 8-bit T1 fonts
\usepackage{hyperref}       % hyperlinks
\usepackage{url}            % simple URL typesetting
\usepackage{booktabs}       % professional-quality tables
\usepackage{amsfonts}       % blackboard math symbols
\usepackage{nicefrac}       % compact symbols for 1/2, etc.
\usepackage{microtype}      % microtypography
\usepackage{lipsum}
\usepackage{graphicx}
\usepackage{tabularx}
\usepackage{listings}
\usepackage{url}
\usepackage{array}
\usepackage{float}
\usepackage{amsmath} 
\usepackage{ragged2e}
\usepackage{placeins}
\graphicspath{ {./images/} }

\title{Integrating Large Language Models for Automated Structural Analysis}

\author{
 Haoran Liang \\
  Department of Civil and Environmental Engineering\\
  University of Alberta\\
  Edmonton, AB T6G 2R3 \\
  \texttt{hliang7@ualberta.ca} \\
   \And
 Mohammad Talebi Kalaleh \\
  Department of Civil and Environmental Engineering\\
  University of Alberta\\
  Edmonton, AB T6G 2R3 \\
  \texttt{talebika@ualberta.ca} \\
  \And
 Qipei Mei \\
  Department of Civil and Environmental Engineering\\
  University of Alberta\\
  Edmonton, AB T6G 2R3 \\
  \texttt{qipei.mei@ualberta.ca} \\
}

\begin{document}
\maketitle
\begin{abstract}

Automated analysis for engineering structures offers considerable potential for boosting efficiency by minimizing repetitive tasks. Although AI-driven methods are increasingly common, no systematic framework yet leverages Large Language Models (LLMs) for automatic structural analysis. To address this gap, we propose a novel framework that integrates LLMs with structural analysis software. LLMs serve as the core engine: they parse structural descriptions from text and translate them into executable Python scripts. Moreover, the framework integrates the generative capabilities of LLMs with code-based finite element (FE) tools like OpenSeesPy. It employs domain-specific prompt design and in-context learning strategies to enhance the LLM's problem-solving capabilities and generative stability, enabling fully automated structural analysis from descriptive text to model outputs. In our experiments, we introduce a well-curated small-scale benchmark dataset of 20 structural analysis word problems (SAWPs) with ground-truth solutions and evaluate the performance of different LLMs within our framework in solving these SAWPs. The role of system instructions, crafted by structural engineers, is also investigated to understand their impact on LLM-driven structural analysis. Additionally, the generative stability of our framework is examined. Through multiple validation experiments on the benchmark, our results demonstrate that the proposed framework can substantially increase the level of automation in solving SAWPs compared to traditional methods. Quantitatively, the framework, built on GPT-4o, achieved 100\% accuracy, surpassing GPT-4 (85\%), Gemini 1.5 Pro (80\%), and Llama-3.3 (30\%) on the test examples. Furthermore, integrating domain-specific instructions enhanced performance by 30\% on problems with asymmetrical structural configurations.

\end{abstract}

\section{Introduction}
Structural engineering depends on thorough analysis to ensure accurate design under various loading conditions. While this analysis can be conducted through hand calculations, finite element modeling (FEM) software, or a combination of both, computational approaches are increasingly favored for their efficiency and ability to handle complex structural systems. These tools range from open-source platforms like OpenSees \cite{mckenna2011opensees}, Code\_Aster \cite{antonutti2018dynamic}, and Calculix \cite{dhondt2017calculix}, to commercial programs such as SAP2000 \cite{pasticier2008non}, Abaqus/CAE \cite{wang2018three}, and ANSYS \cite{kohnke1982ansys}. They support a wide spectrum of structures, from 2D frames \cite{elkady2021open} and trusses \cite{radfar2017finite} to high-rise buildings \cite{verma2023evaluation} and bridges \cite{hejazi2021solving}. However, traditional FEM-based methods still require substantial manual effort and specialized knowledge, highlighting the need for further automation and optimization in structural analysis.

Large Language Models (LLMs) offer a promising opportunity to revolutionize different sectors by automating laborious tasks while preserving engineering rigor. Recent advances in transformer architectures \cite{vsp+17} have demonstrated remarkable capabilities not only in natural language processing (NLP) \cite{raffel2020exploring,brown2020language,wei2022chain} but also in computer vision (CV) \cite{peebles2023scalable, khan2022transformers,ssz+25_dit}. Moreover, these models have broadened their impact by addressing diverse language-related challenges in scientific fields \cite{taylor2022galactica,khan2023chatgpt,hendrycks2020aligning}. The rapid emergence of LLMs has become a defining trend in modern NLP research, marked by key milestones such as the development of the base model GPT-3 \cite{brown2020language}, the instruction-tuned model InstructGPT \cite{ouyang2022training}, the advanced multi-modal and fine-tuned model GPT-4 \cite{achiam2023gpt}, reinforcement learning-enhanced reasoning models such as DeepSeek-R1 \cite{guo2025deepseek} and OpenAI’s model o1 \cite{jaech2024openai}, as well as leading open-source models like Llama \cite{touvron2023llama} and Qwen \cite{bai2023qwen}.  Despite the immense potential of LLMs, most current research focuses primarily on their performance in general NLP tasks \cite{chen2021capturing, chen2022scientific}.
However, beyond these task-specific evaluations, several studies have explored strategies to optimize LLMs for broader applicability and efficiency. For instance, pruning techniques \cite{lls+24_prune, ssz+25_prune} have been investigated to improve inference speed and reduce memory usage, while differential privacy methods \cite{lssz24_dp} have been applied to enhance data security and protect user confidentiality without compromising model utility.

Despite the significant advances in LLM-driven applications across various practical domains—such as healthcare \cite{schmidgall2024agentclinic}, finance \cite{xiao2024tradingagents}, and data science \cite{lei2024spider}, civil engineering has received comparatively less attention. Recently, researchers have begun exploring their applications in civil engineering, including LLM-based architectural flaw detection \cite{kumar2024architectural} and LLM-assisted technical writing for urban construction \cite{cruz2024llm}. Another study \cite{fan2024ergochat} introduced an interactive visual query system designed to assess the ergonomic postural risks of construction workers. This system incorporates visual question answering (VQA) to respond to visual queries regarding workers' exposure to ergonomic risks and image captioning (IC) to generate textual descriptions of these risks from images. Furthermore, a study \cite{Joffe2025} proposed an LLM-based tool that enables engineers to ask code-related questions in natural language and receive accurate answers with citations, demonstrating its effectiveness using the 2020 National Building Code of Canada and highlighting its potential to improve design efficiency.

More recently, there has been some initial work exploring LLM's application in structural engineering, particularly structural analysis and design \cite{orlando2023assessing}. Researchers have used ChatGPT to generate code for solving a Poisson equation, utilizing the Python interfaces of deal.II and FEniCS. Essentially, they prompted ChatGPT to produce the necessary code for addressing the problem. Further work examined interactions between multiple LLM-based agents for programming tasks, utilizing FEniCS for FEA and GPT-3-turbo for code generation \cite{tian2024optimizing}. The researchers applied this setup to a simple 2D plate structure, yet the method faces scalability issues. With only one basic example, comparing different agents’ performances becomes less persuasive and informative. Similarly, research on AI-driven design optimization has gained attention. A study \cite{qin2024intelligent} proposes an intelligent design and optimization system for shear wall structures, leveraging LLMs and generative artificial intelligence. This system employs an LLM as the central controller, interpreting engineers' language descriptions and converting them into executable code. Despite growing interest in this field, there are two critical challenges that stand in the way of realizing the full potential of LLMs in structural analysis: 1) Determining the potential capabilities of LLMs in structural analysis requires a systematic analysis of both the models themselves and the specific demands of structural engineering tasks. 2) Conducting a reliable and comprehensive evaluation necessitates diverse experimental settings and careful consideration of factors such as standardized evaluation procedures, dataset curation, prompt design, and in-context learning strategies.

To address these gaps, we propose a novel framework that combines the generative capabilities of LLMs with the OpenSeesPy package \cite{zhu2018openseespy}, and we assess its performance on a curated dataset of 20  structural analysis word problems (SAWPs). Specifically, we employ multiple base models within the framework—including GPT-4 \cite{achiam2023gpt}, GPT-4o \cite{hurst2024gpt}, Llama 3 \cite{dubey2024llama}, and Gemini 1.5 \cite{team2024gemini}—and compare their baseline performance with versions enhanced by techniques such as few-shot learning \cite{brown2020language} and in-context learning (ICL) \cite{dong2024survey}. LLMs are capable of extracting critical information from textual problem descriptions and generating Finite Element Analysis (FEA) Python scripts for 2D frame structures. This enables engineers to provide concise natural language prompts rather than manually debugging code or interacting directly with analysis software, substantially reducing the time required for structural modeling and results visualization.

The remainder of this paper is structured as follows. In Section \ref{methodology}, we detail the core methodology, including the tools and techniques employed by LLMs to solve SAWPs. Section \ref{experimentsetup} outlines the computational setup and the design of our 20-problem benchmark. Section \ref{results} presents key findings on a comparative performance analysis, generative stability and the influence of system instructions on output quality. Finally, Section \ref{conclusion} summarizes the main conclusions and outlines potential directions for future research.

\FloatBarrier
\section{Methodology}\label{methodology}

\subsection{General Workflow}
\begin{figure}[h]
    \centering
    \includegraphics[width=1\textwidth, trim=50 100 450 0, clip]{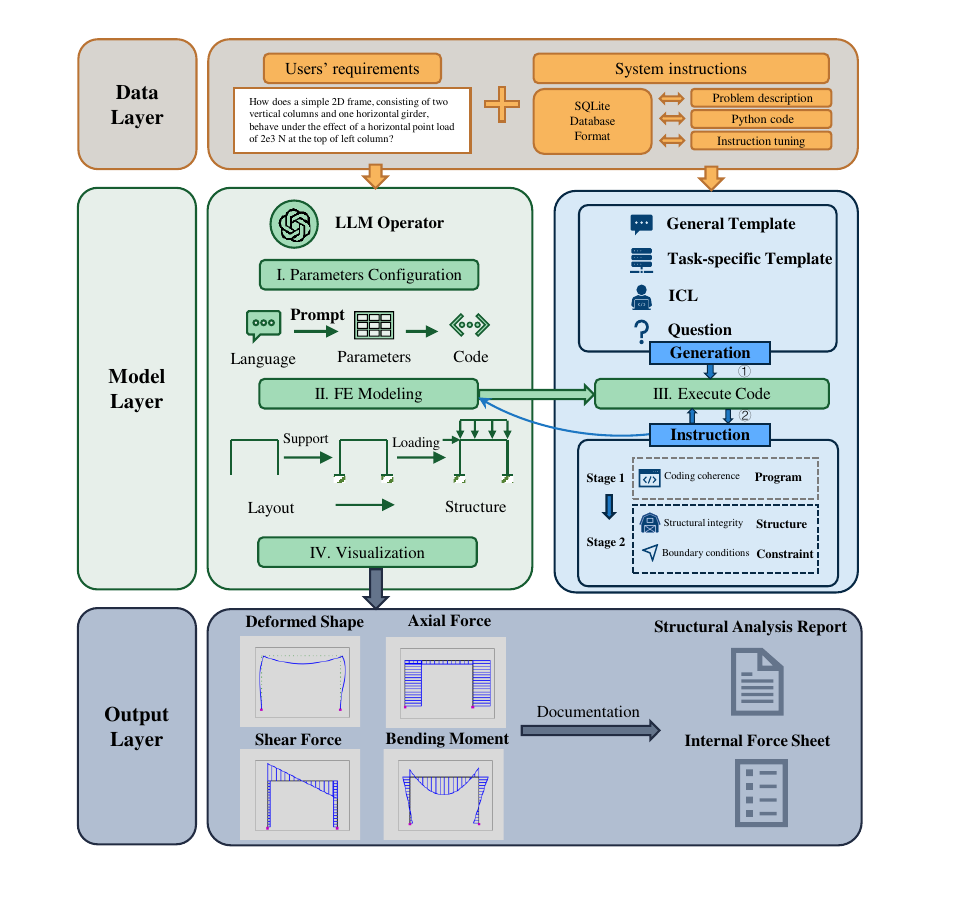}  
    \caption{LLM-Driven finite element analysis framework for 2D frame structures}
    \label{workflow}
\end{figure}

The workflow of our model is illustrated in Figure \ref{workflow}. In Section \ref{data layer}, we described the user requirements, the construction of system instructions, and the data structures used to store them. In Section \ref{model layer}, we detailed how structural analysis problems are decomposed, how the LLM processes each component, and how system instructions enhance the model’s problem-solving capabilities within our framework. Finally, in Section \ref{output layer}, we discussed the types of visualizations generated by our framework and the output format.

\subsection{Data Layer}\label{data layer}
At the data layer, the input to the LLM comprises two components: (1) the user's requirements, provided as a problem description, and (2) system instructions, which integrate standard few-shot prompting, as introduced by Brown et al. \cite{brown2020language}. And our framework also introduces system instructions to enhance the LLM’s ability to understand and solve SAWPs. Specifically, we use SQLite to structure the system instructions, which include: the problem description (formatted similarly to the user's requirements), the corresponding Python code for performing structural analysis and visualization as requested in the problem description (crucial for in-context learning, allowing the LLM to learn and replicate effective problem-solving formats), and instruction tuning elements—such as intermediate reasoning steps—to guide the LLM in solving specific sub-tasks. 

\subsection{Model Layer}\label{model layer}

At the model layer, the LLM serves as the primary operator. We use the OpenSeesPy package in Python for structural analysis. OpenSeesPy is the Python interface for OpenSees, an open-source finite element analysis software specifically designed for structural and earthquake engineering simulations. It supports 2D frame analysis through Python scripting, enabling the LLM to generate and execute Python code to produce outputs such as complete analysis scripts and internal force diagrams. Structural analysis involves extensive numerical computation, and it is considerably more reliable to utilize an LLM's tool-use capabilities—such as generating Python code—rather than depending solely on its probabilistic recall of information from Internet documents. To enhance reliability during the generation stage and support effective debugging, we divide the code generation into three distinct stages. Specifically, our framework invokes the LLM through API (Application Programming Interface) three times—once for each stage of the code generation process per problem.  The first stage extracts key parameters—such as material and geometric properties—from the problem description. The second and most complex stage involves generating the structural layout, assigning supports, and defining loading conditions. The third stage focuses on visualization, where the LLM generates diagrams in accordance with the user’s specifications. This structured workflow mitigates errors often caused by overly long code generations (a common issue when prompting LLMs to produce lengthy solutions in a single pass), improves execution success rates, and simplifies the identification of failures by isolating them within specific code segments. Furthermore, the word problem descriptions were formulated using a standardized format to ensure consistency within the benchmark. This consistency enhances the reliability of the evaluation results and enables clearer observation of our framework’s characteristics through comparison when running it on different examples from the benchmark. We also adjust the system instructions to test whether refinements suggested by experienced structural engineers can enhance the model’s accuracy and reliability. An example of how the LLM extracts information from a problem description and converts it into Python scripts is shown in Figure \ref{methodology}.

\begin{figure}[h]
    \centering
    \includegraphics[width=1\textwidth, trim=50 0 50 0, clip]{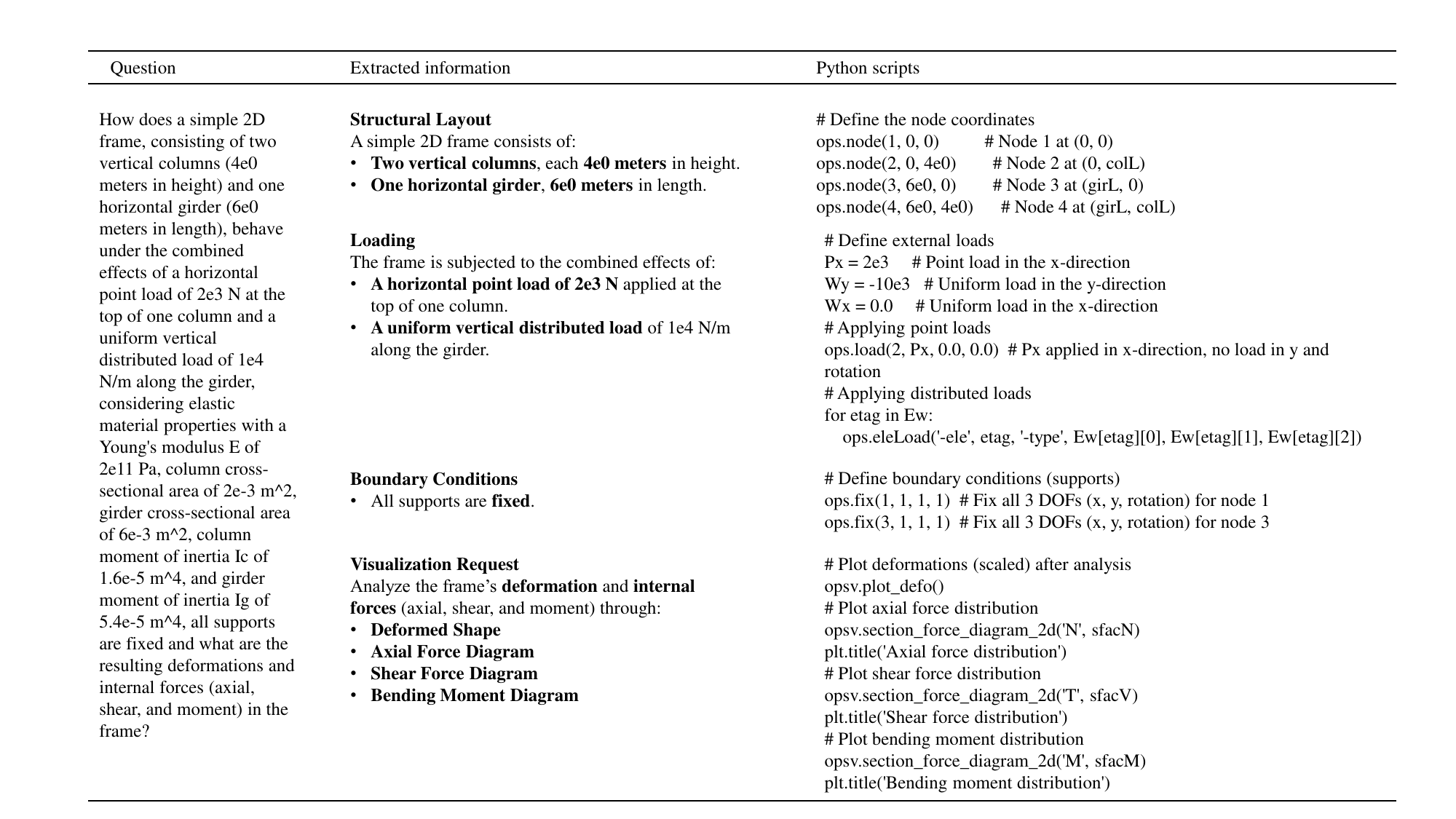}  
    \caption{Partial process illustrating how LLMs extract information from the problem description and convert it into Python scripts.}
    \label{methodology}
\end{figure}

\FloatBarrier
\textbf{Toolkit.} LLMs themselves cannot perform precise structural analysis. However, their ability to generate code, particularly in Python, has significantly improved \cite{chen2021evaluating}. To leverage this capability, we employ LLMs to interface with two Python libraries: OpenSeesPy, a popular open-source Python package for finite element structural analysis, and OpsVis, a specialized library for visualizing results obtained by OpenSeesPy. Our instruction set for the LLM includes a step-by-step guide on using these libraries to solve a basic 2D frame structural analysis problem. The LLM is responsible for both performing the coding for structural analysis and accomplish visualizations using these tools. Traditionally, solving such problems with these libraries requires expertise in structural mechanics and Python programming. But by utilizing the LLM as a compiler, we enable a direct translation of natural language problem descriptions into desired results, eliminating the need for extensive domain-specific knowledge and coding skills.

\begin{figure}[h]
    \centering
    \includegraphics[width=1\textwidth]{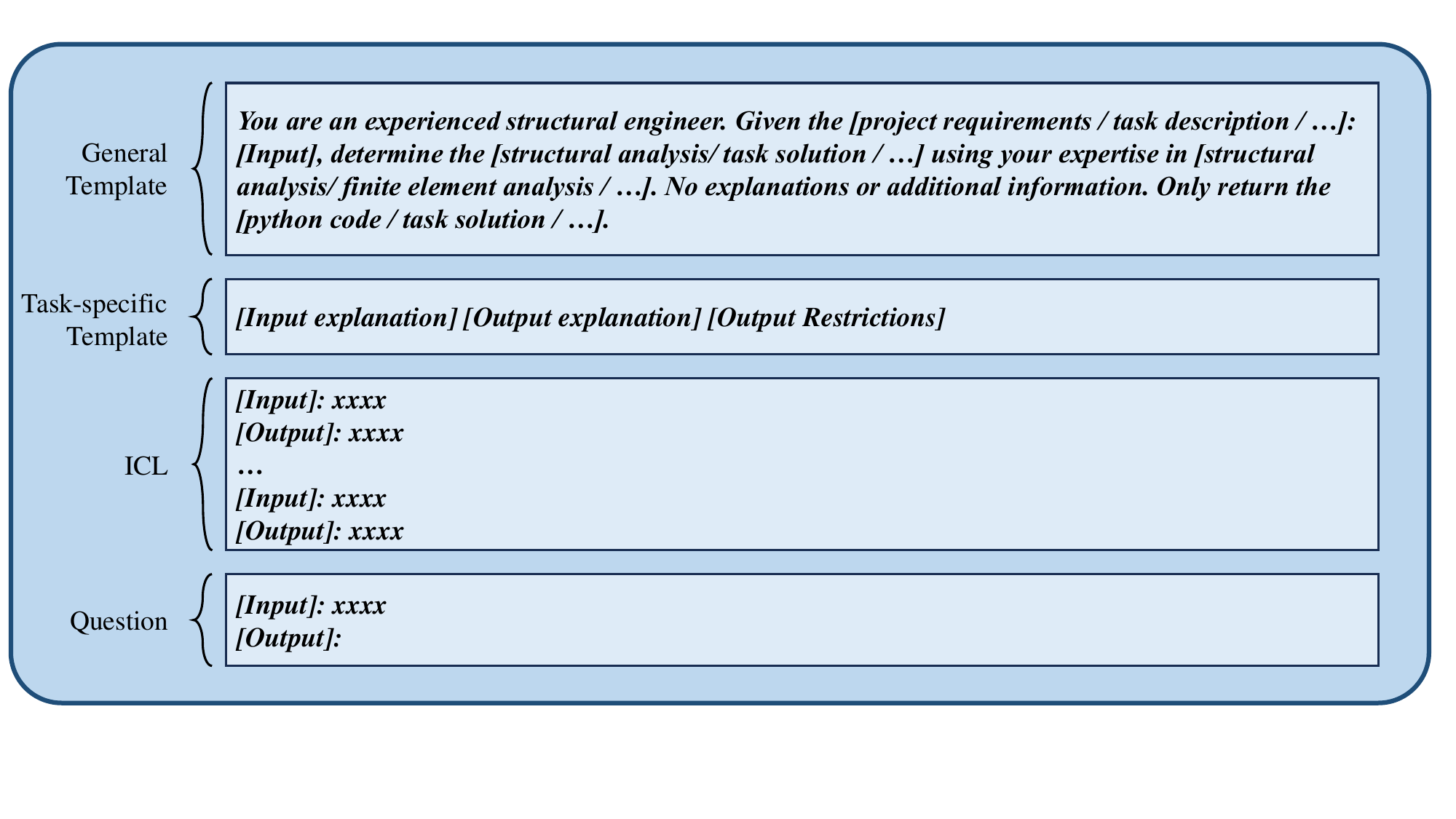}  
    \caption{An ICL prompt template for all structural analysis problems}
    \label{ICL}
\end{figure}

\FloatBarrier
\textbf{ICL Prompt Template.} To enhance the LLM’s understanding of structural engineering problems, we provide more than just the problem description. Our input includes an example problem, its corresponding code, and specific constraints to help the LLM generate runnable and stable Python scripts. To structure the system instructions effectively, we introduce an ICL prompt template in Figure \ref{ICL}. ICL has been shown to improve LLM performance in coding and reasoning \cite{dong2024survey}, and the style and format of our template are adapted from reference \cite{guo2023can}. \textbf{General Template} includes preliminary settings to help the LLM understand the problem in a structural engineering context. \textbf{Task-Specific Template} provides explanations and constraints for both input and output. \textbf{ICL} contains a reference problem, using Example 1 from Table \ref{schematic_partial}, along with detailed instructions on writing Python scripts using OpenSeesPy and OpsVis. Finally, \textbf{Question} presents the problem that the LLM needs to solve. The complete ICL template is shown in \ref{icltemplate}.

\begin{figure}[h]
    \centering
    \includegraphics[width=1\textwidth]{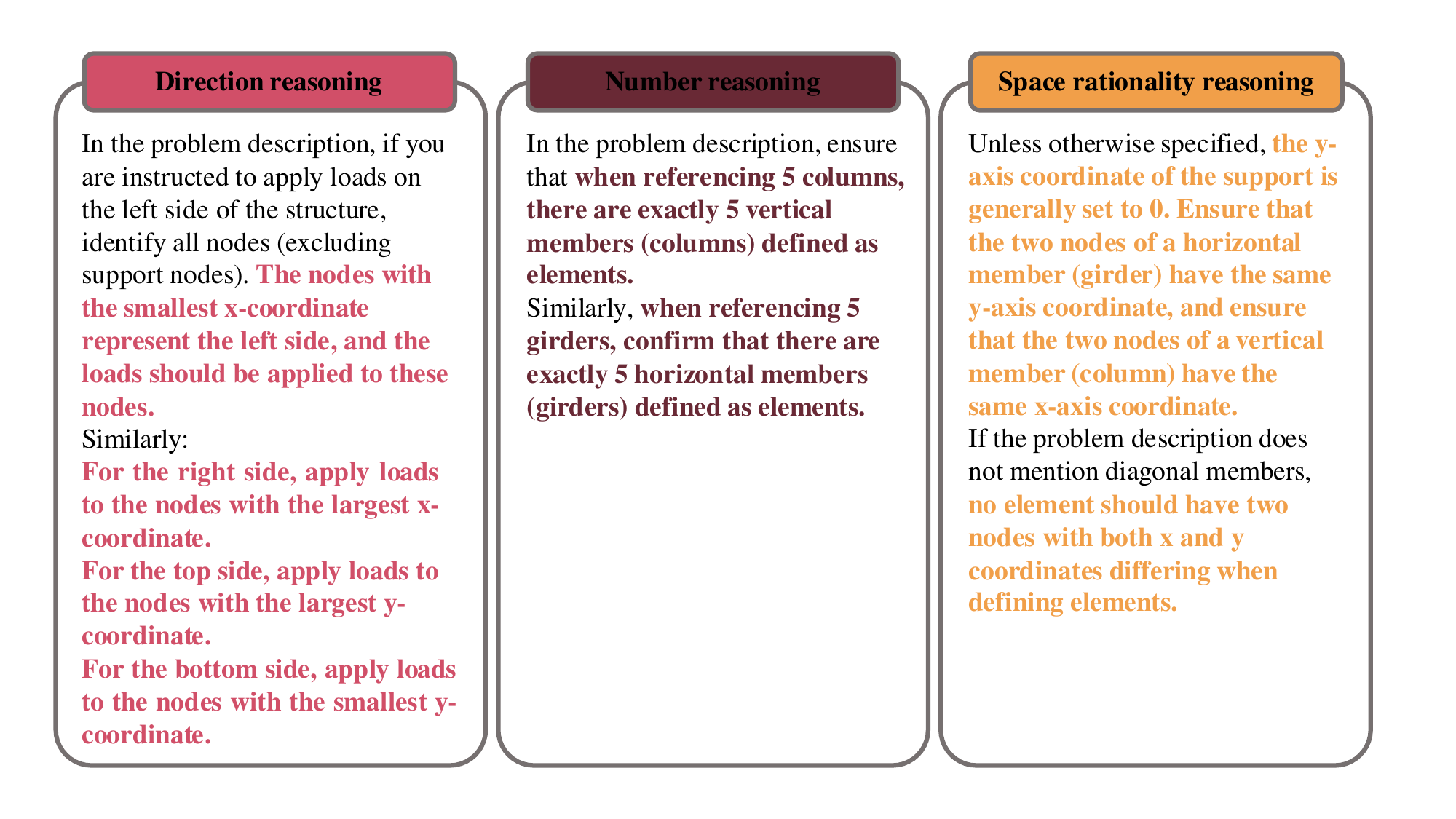}  
    \caption{Examples of commonsense reasoning for system instructions}
    \label{commonsense}
\end{figure}

\textbf{Commonsense Reasoning.} Instruction tuning has been shown to improve LLM performance across various reasoning tasks \cite{wei2022chain}. Since structural analysis problems require LLMs to interpret problem descriptions and reason about structural layouts, we incorporate essential commonsense reasoning into the system instructions to enhance problem-solving capabilities. Examples of commonsense reasoning used in our approach are shown in Figure\ref{commonsense}. These examples are for illustration purposes only; the distributed loading direction reasoning is presented in Table\ref{full_instructions}.

\FloatBarrier
\subsection{Output Layer}\label{output layer}
At the output layer, the results generated by the model layer are collected and formatted into structured outputs, such as reports or spreadsheets, depending on the user’s requirements.  To be more specific, with just a few lines of prompt, the LLM is able to automatically generate a structured report that includes the problem description, the generated code, explanations of the results, and spreadsheets containing internal forces at each node, along with their maximum and minimum values. By systematically processing inputs, integrating structured instructions, and producing clear, visualized outputs, our framework significantly enhances the capability of LLMs to solve SAWPs, while maintaining transparency and interpretability throughout the generative process.
\FloatBarrier
\section{Experiment Setup}\label{experimentsetup}

\subsection{Computational Setup}

We used several LLMs in this work, including gpt-4o-2024-08-06 and gpt-4-turbo from OpenAI, Gemini 1.5 Pro from Google, and llama-3.3-70b-versatile from Meta. These models were accessed via API calls, minimizing the need for local computational resources. Most API-based tasks completed within a few seconds per request. All scripts were written in Python 3.12 and executed on a Lenovo Legion 5 (2022) laptop running Windows 11. The machine is equipped with a Ryzen 7 6800H CPU (8 cores, up to 4.7\,GHz), 64\,GB RAM, 1\,TB SSD, and an NVIDIA GeForce RTX 3060 GPU. The key Python libraries used to call LLMs API include \texttt{openai}, \texttt{google-generativeai}, and \texttt{groq}. API keys were securely stored using environment variables.

\subsection{Test Examples}
The paper focuses on a foundational structural pattern: simple 2D frames. These structures are straightforward to describe within a single dialogue turn. Currently, there is no publicly available dataset that systematically includes structural descriptions, clear layouts, and corresponding ground truth. To fill this gap, we manually designed a benchmark comprising 20 different structural analysis problems. For each structure, we provided a detailed word problem description along with ground truth solutions to validate the outputs generated by LLMs. Several key problems appear in Table \ref{schematic_partial}, while the full set of problems is included in \ref{dataset}.

\begin{table}[h]
    \centering
    \renewcommand{\arraystretch}{1.5}
    \caption{Partial problem descriptions and ground truth schematics}
    \label{schematic_partial}
    \setlength{\extrarowheight}{2pt}
    \renewcommand{\tabularxcolumn}[1]{>{\hsize=\dimexpr\hsize+2\tabcolsep\relax\justifying\arraybackslash}m{#1}}
    \begin{tabularx}{\textwidth}{X >{\centering\arraybackslash}m{0.32\textwidth}} 
        \toprule
        \textbf{Problem description} & \textbf{Ground truth} \\
        \midrule
        \textbf{1. }How does a simple 2D frame, consisting of two vertical columns (\(4 \times 10^0\) meters in height) and one horizontal girder (\(6 \times 10^0\) meters in length), behave under the combined effects of a horizontal point load of \(2 \times 10^3\) N at the top of one column and a uniform vertical distributed load of \(1 \times 10^4\) N/m along the girder, considering elastic material properties with a Young's modulus \(E = 2 \times 10^{11}\) Pa, column cross-sectional area \(A_c = 2 \times 10^{-3} \, \text{m}^2\), girder cross-sectional area \(A_g = 6 \times 10^{-3} \, \text{m}^2\), column moment of inertia \(I_c = 1.6 \times 10^{-5} \, \text{m}^4\), and girder moment of inertia \(I_g = 5.4 \times 10^{-5} \, \text{m}^4\). All supports are fixed. What are the resulting deformations and internal forces (axial, shear, and moment) in the frame?
        & \begin{minipage}[t][4cm][t]{0.32\textwidth}
            \centering
            \includegraphics[scale=0.5]{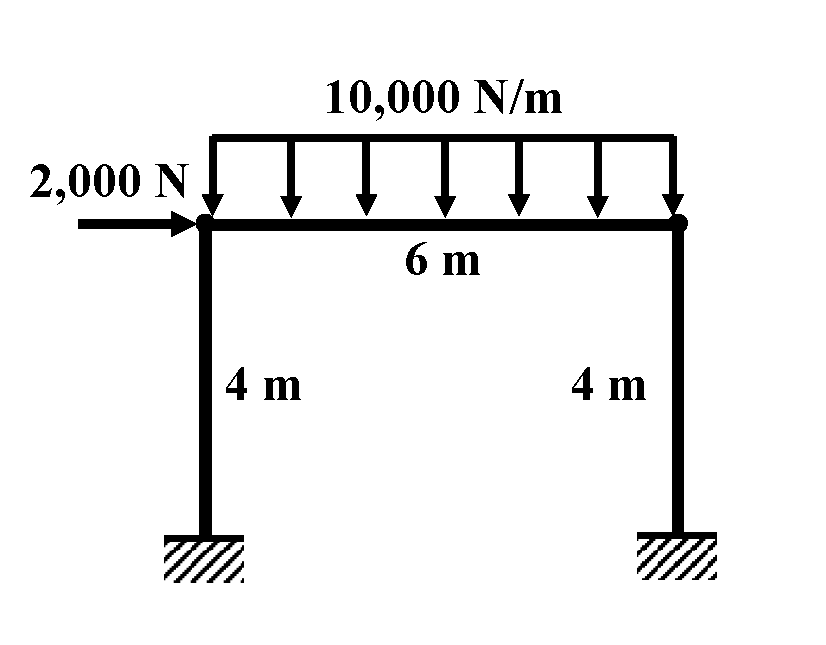}
        \end{minipage} \\
        \midrule
        \textbf{12. }How does a two-story two-bay 2D frame, where the first bay has two stories and the second bay has one story, consisting of 5 vertical columns (\(4\) meters in height each) and 3 horizontal girders (\(6\) meters in length each), behave under the uniform vertical distributed load of \(1 \times 10^4\) N/m along each girder? Consider elastic material properties with Young's modulus of \(2 \times 10^{11}\) Pa, column cross-sectional area of \(2 \times 10^{-3}\) m\(^2\), girder cross-sectional area of \(6 \times 10^{-3}\) m\(^2\), column moment of inertia of \(1.6 \times 10^{-5}\) m\(^4\), and girder moment of inertia of \(5.4 \times 10^{-5}\) m\(^4\). All supports are fixed. What are the resulting deformations and internal forces (axial force, shear force, and bending moment) within the frame?
        & \begin{minipage}[t][4cm][t]{0.32\textwidth}
            \centering
            \includegraphics[scale=0.35]{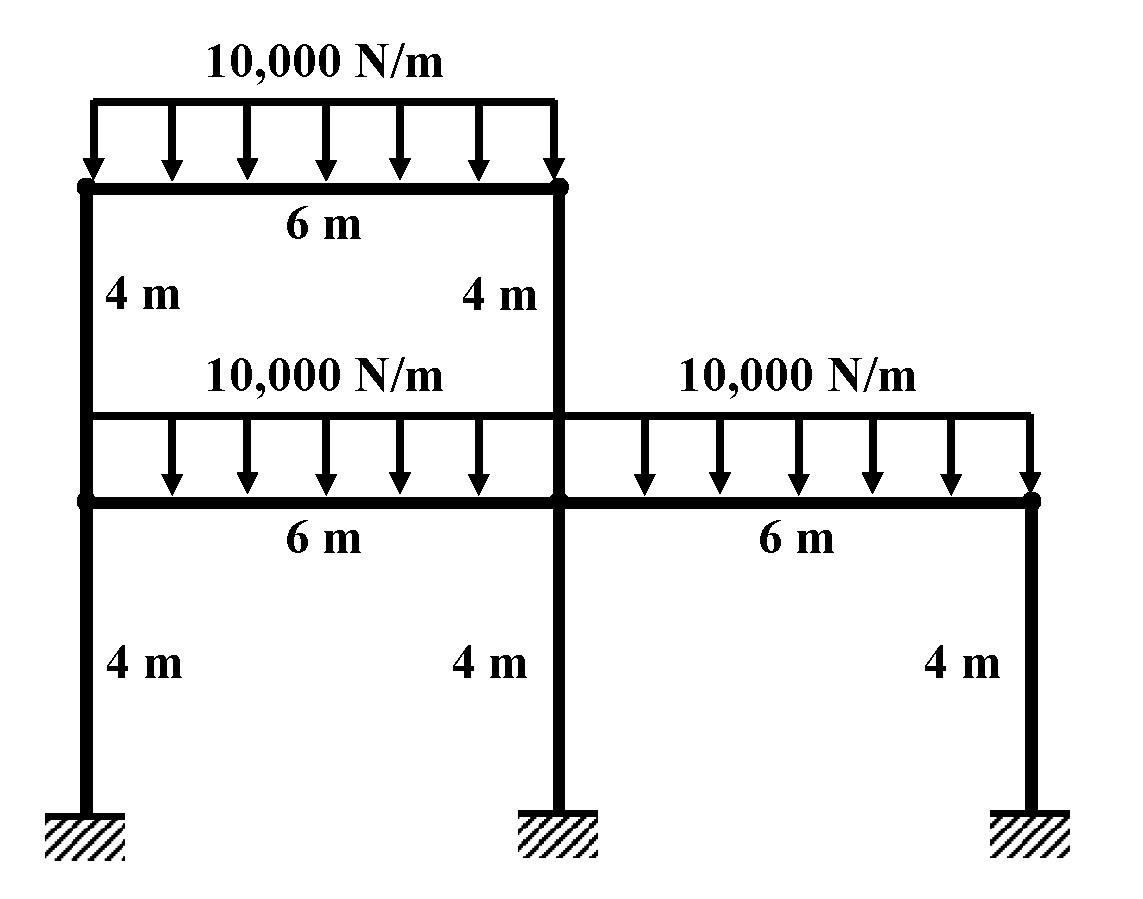}
        \end{minipage} \\
        \midrule
        \textbf{20. }How does a simple 2D frame, consisting of two vertical columns (\(4 \times 10^0\) meters in height), one horizontal girder (\(6 \times 10^0\) meters in length) and two cantilever beams (\(2 \times 10^0\) meters in length) on both sides which are connected to the top of two columns, behave under the uniform vertical distributed load of (\(1 \times 10^4\)) N/m along the girder and two cantilever beams? Consider elastic material properties with a Young's modulus \(E\) of (\(2 \times 10^{11}\)) Pa, column cross-sectional area of (\(2 \times 10^{-3}\)) m\(^2\), girder and cantilever beam cross-sectional area of (\(6 \times 10^{-3}\)) m\(^2\), column moment of inertia of (\(1.6 \times 10^{-5}\)) m\(^4\), and girder and cantilever beam moment of inertia of (\(5.4 \times 10^{-5}\)) m\(^4\). All supports are fixed. What are the resulting deformations and internal forces (axial, shear, and moment) in the frame?
        & \begin{minipage}[t][4.5cm][t]{0.32\textwidth}
            \centering
            \includegraphics[scale=0.5]{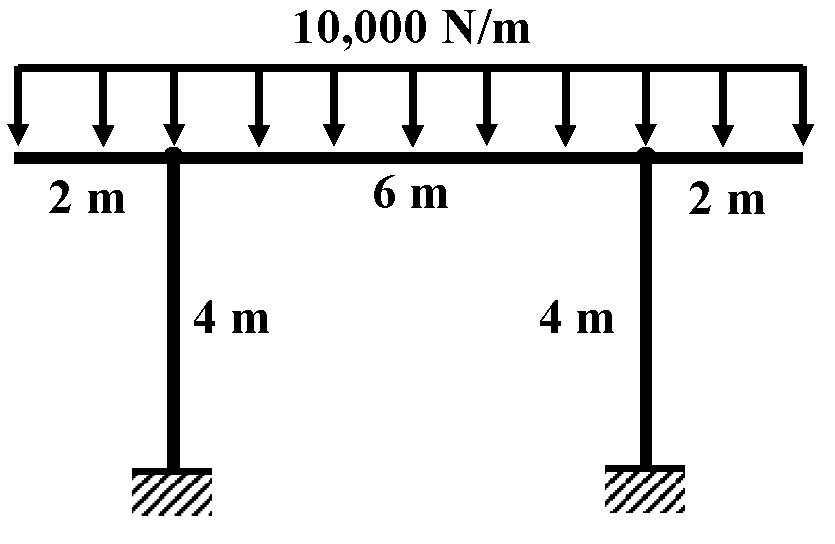}
        \end{minipage} \\
        \bottomrule
    \end{tabularx}
\end{table}

\begin{figure}[h]
    \centering
    \begin{subfigure}[b]{0.24\textwidth}
        \centering
        \includegraphics[width=\textwidth]{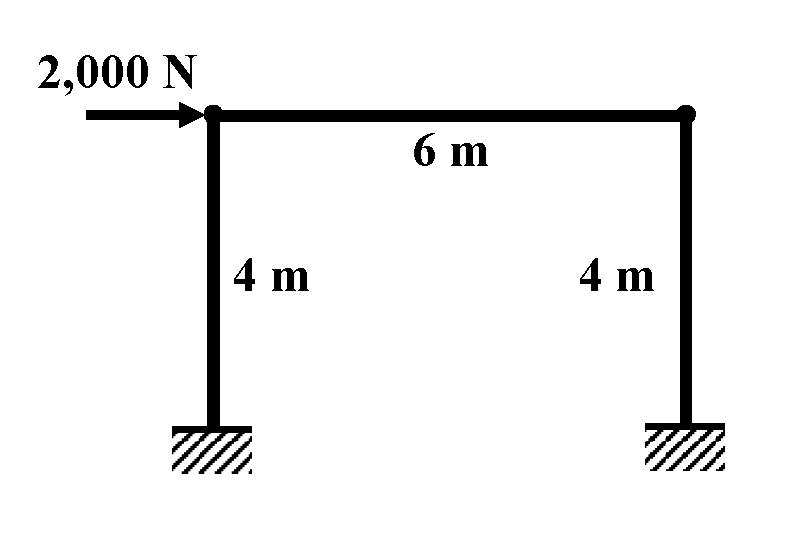}
        \caption{Example 3}
        \label{benchmark1_3}
    \end{subfigure}
    \hfill
    \begin{subfigure}[b]{0.24\textwidth}
        \centering
        \includegraphics[width=\textwidth]{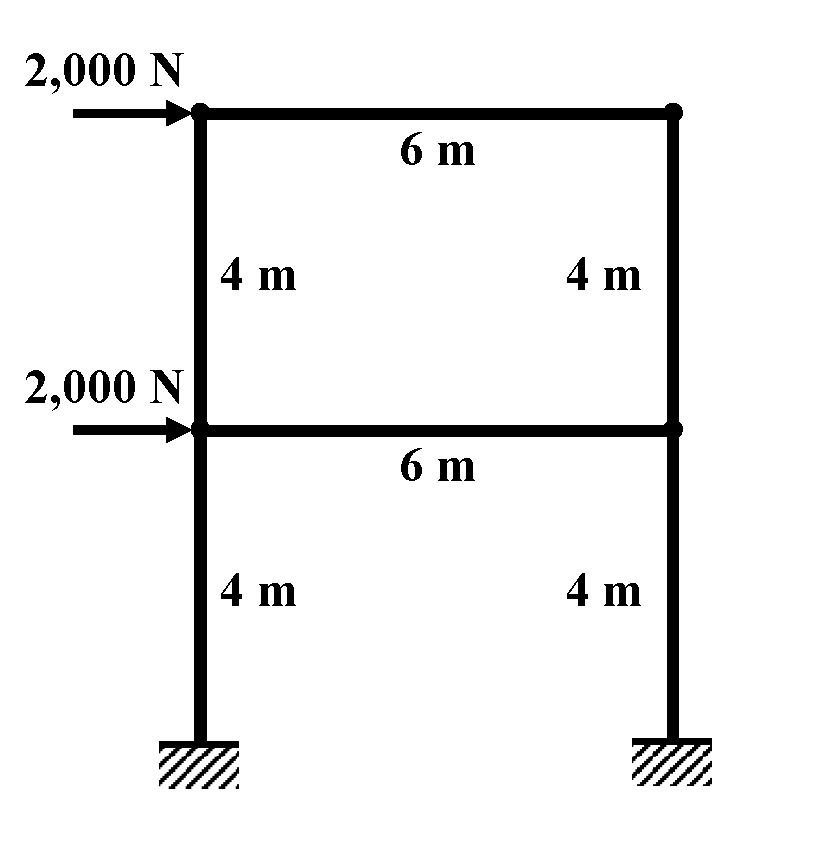}
        \caption{Example 5}
        \label{benchmark1_5}
    \end{subfigure}
    \hfill
    \begin{subfigure}[b]{0.48\textwidth}
        \centering
        \includegraphics[width=\textwidth]{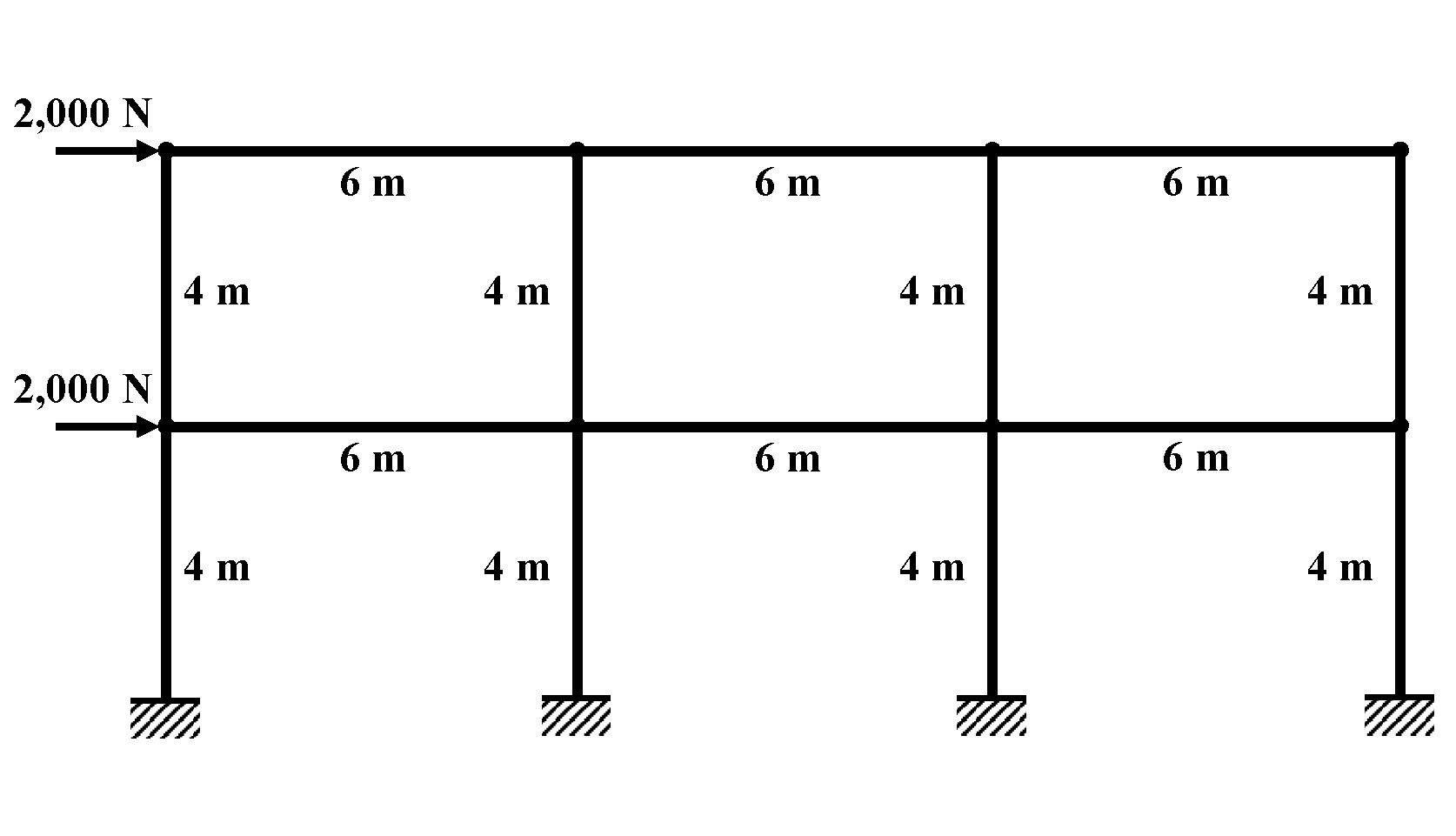}
        \caption{Example 9}
        \label{benchmark1_9}
    \end{subfigure}
    \caption{Pattern 1 for generating new example problems}
    \label{benchmark_359}
\end{figure}

\begin{figure}[h]
    \centering
    \begin{subfigure}[b]{0.25\textwidth}
        \centering
        \includegraphics[width=\textwidth]{images/schematic_05.png}
        \caption{Example 5}
        \label{benchmark2_5}
    \end{subfigure}
    \hfill
    \begin{subfigure}[b]{0.35\textwidth}
        \centering
        \includegraphics[width=\textwidth]{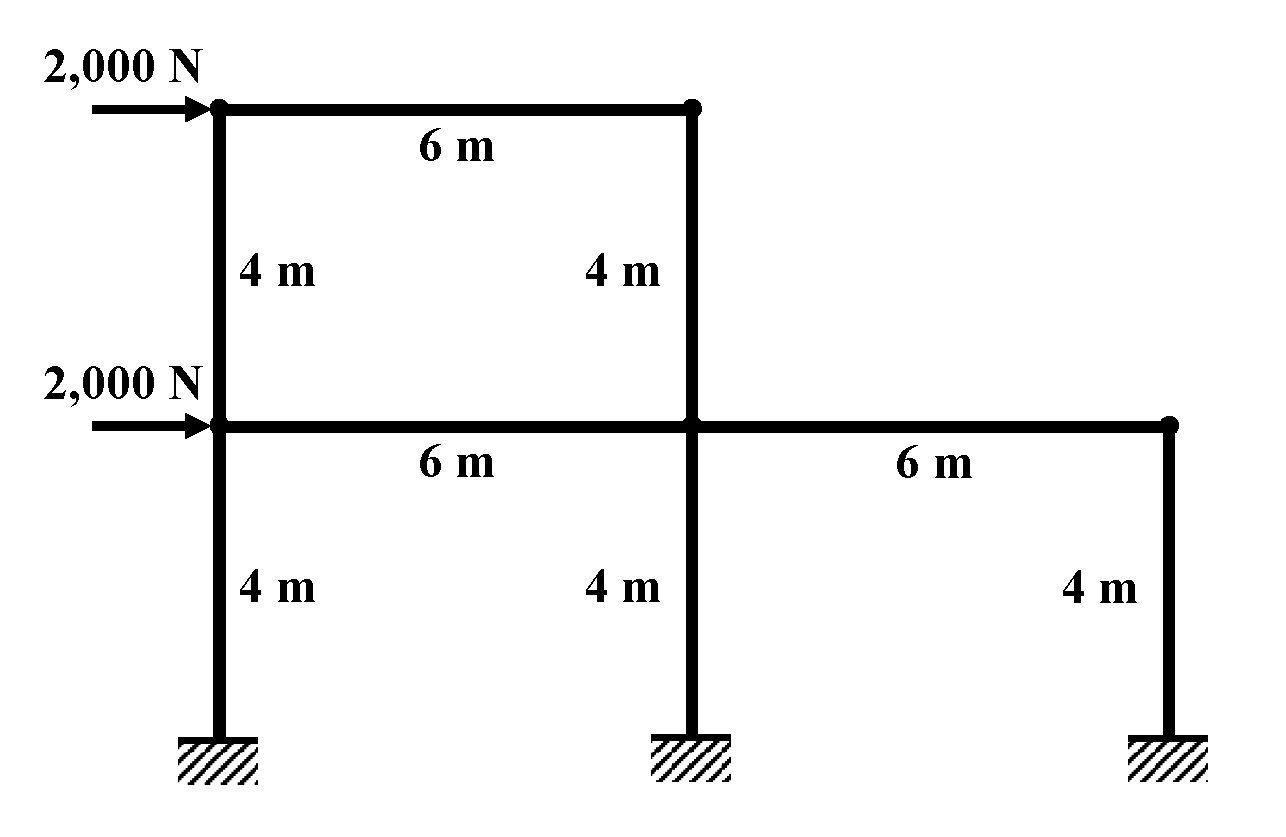}
        \caption{Example 11}
        \label{benchmark2_11}
    \end{subfigure}
    \hfill
    \begin{subfigure}[b]{0.35\textwidth}
        \centering
        \includegraphics[width=\textwidth]{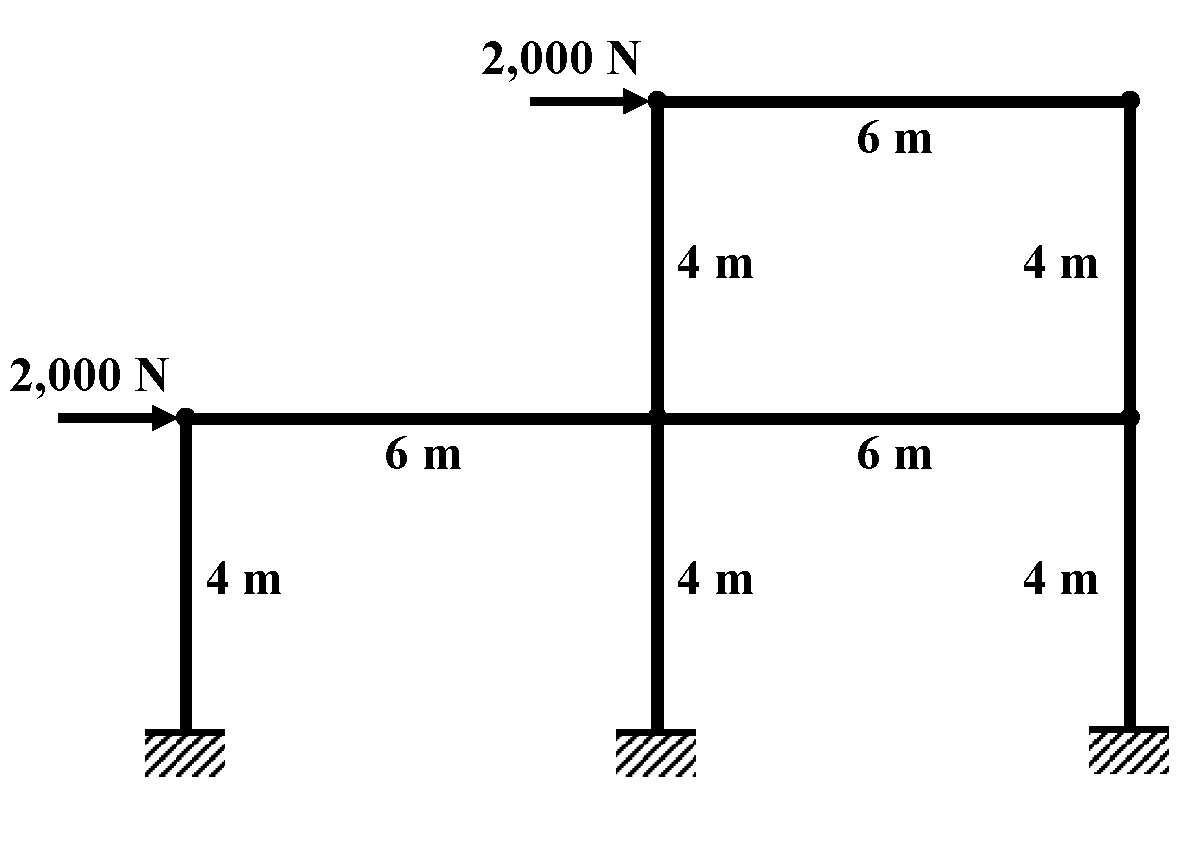}
        \caption{Example 13}
        \label{benchmark2_13}
    \end{subfigure}
    \caption{Pattern 2 for generating new example problems}
    \label{benchmark_51113}
\end{figure}

\begin{figure}[h]
    \centering
    \begin{subfigure}[b]{0.25\textwidth}
        \centering
        \includegraphics[width=\textwidth]{images/schematic_03.png}
        \caption{Example 3}
        \label{benchmark3_3}
    \end{subfigure}
    \hfill
    \begin{subfigure}[b]{0.35\textwidth}
        \centering
        \includegraphics[width=\textwidth]{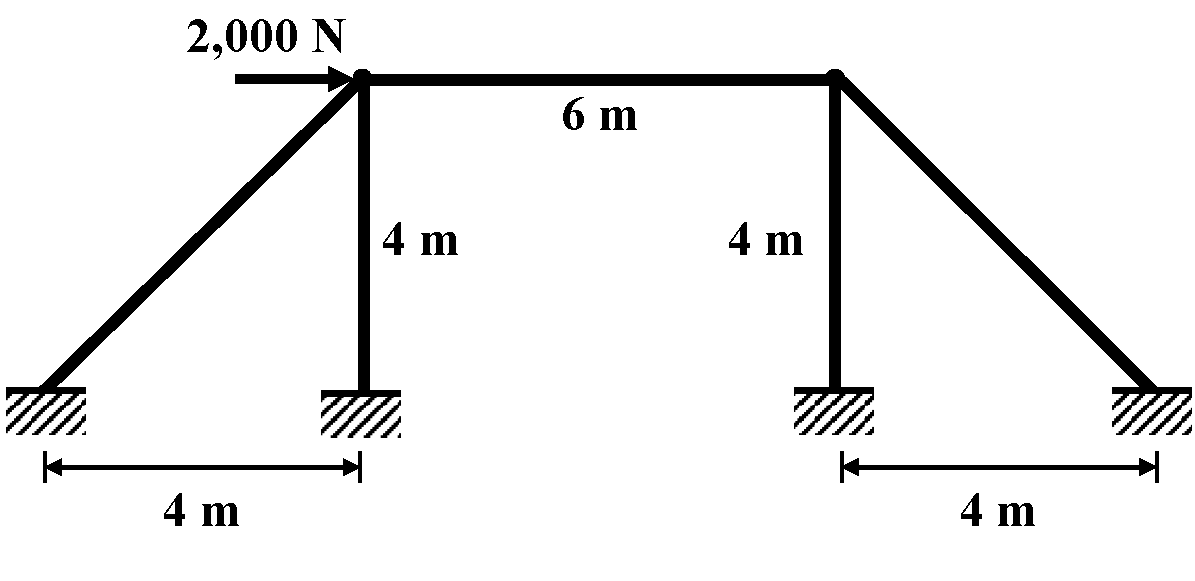}
        \caption{Example 17}
        \label{benchmark3_17}
    \end{subfigure}
    \hfill
    \begin{subfigure}[b]{0.35\textwidth}
        \centering
        \includegraphics[width=\textwidth]{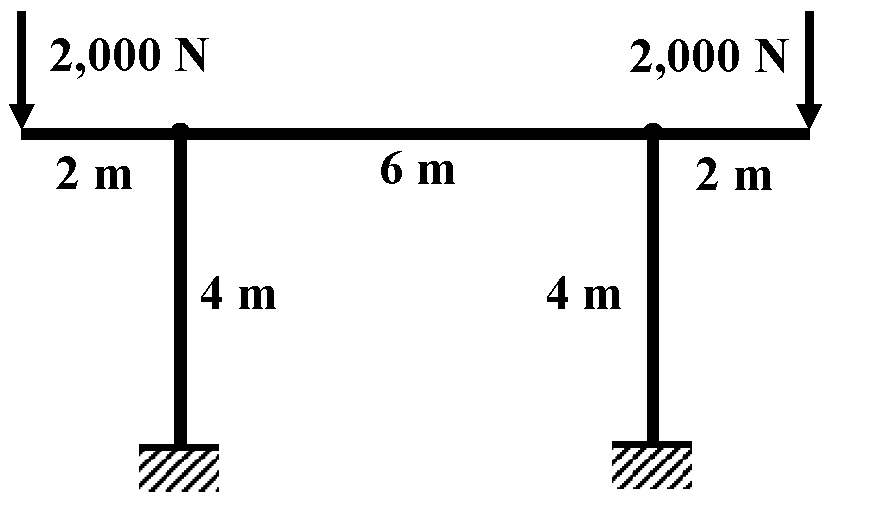}
        \caption{Example 19}
        \label{benchmark3_19}
    \end{subfigure}
    \caption{Pattern 3 for generating new example problems}
    \label{benchmark_31719}
\end{figure}

 To be more specific, each problem was presented in a standardized format, consisting of five main components: geometrical properties, material properties, loading conditions, boundary conditions, and requirements for result visualization. Corresponding schematics were prepared as ground truth for validating the accuracy of the model's outputs. In the creation of the dataset, we used LLMs to generate code based on manually defined problem descriptions, and the generated results are reviewed. This approach significantly reduced the time required compared to manually creating the entire dataset. We adapted three patterns to generate new example problems. In Pattern 1 (Figure \ref{benchmark_359}), we modify the number of stories and bays to create new structures. For instance, transitioning from Figure \ref{benchmark1_3} to Figure \ref{benchmark1_5}, we extend the structure from one to two stories, while from Figure \ref{benchmark1_3} to Figure \ref{benchmark1_9}, we expand it from a single-bay, one-story structure to a three-bay, two-story structure. This pattern requires LLMs to accurately define nodes and elements. In Pattern 2 (Figure \ref{benchmark_51113}), we introduce asymmetry. For example, in Figure \ref{benchmark2_11}, the first bay consists of two stories, while the second bay has only one floor, requiring LLMs to infer and handle structural asymmetry. In Pattern 3, we incorporate additional structural features commonly found in engineering, such as diagonal members (Figure \ref{benchmark3_17}) and cantilever beams (Figure \ref{benchmark3_19}). This pattern challenges LLMs to interpret and solve problems involving complex structural configurations.

\clearpage

\FloatBarrier
\section{Results and Discussion}\label{results}

This section begins with a comparative analysis of our framework across different LLMs' performance, followed by results on the generative stability of our framework on the benchmark and the impact of instructions on output quality.

\subsection{Comparative Analysis of LLM-Generated Structural Analysis}

\begin{figure}[h]
    \centering
    \includegraphics[width=0.6\textwidth,trim=50 0 50 0, clip]{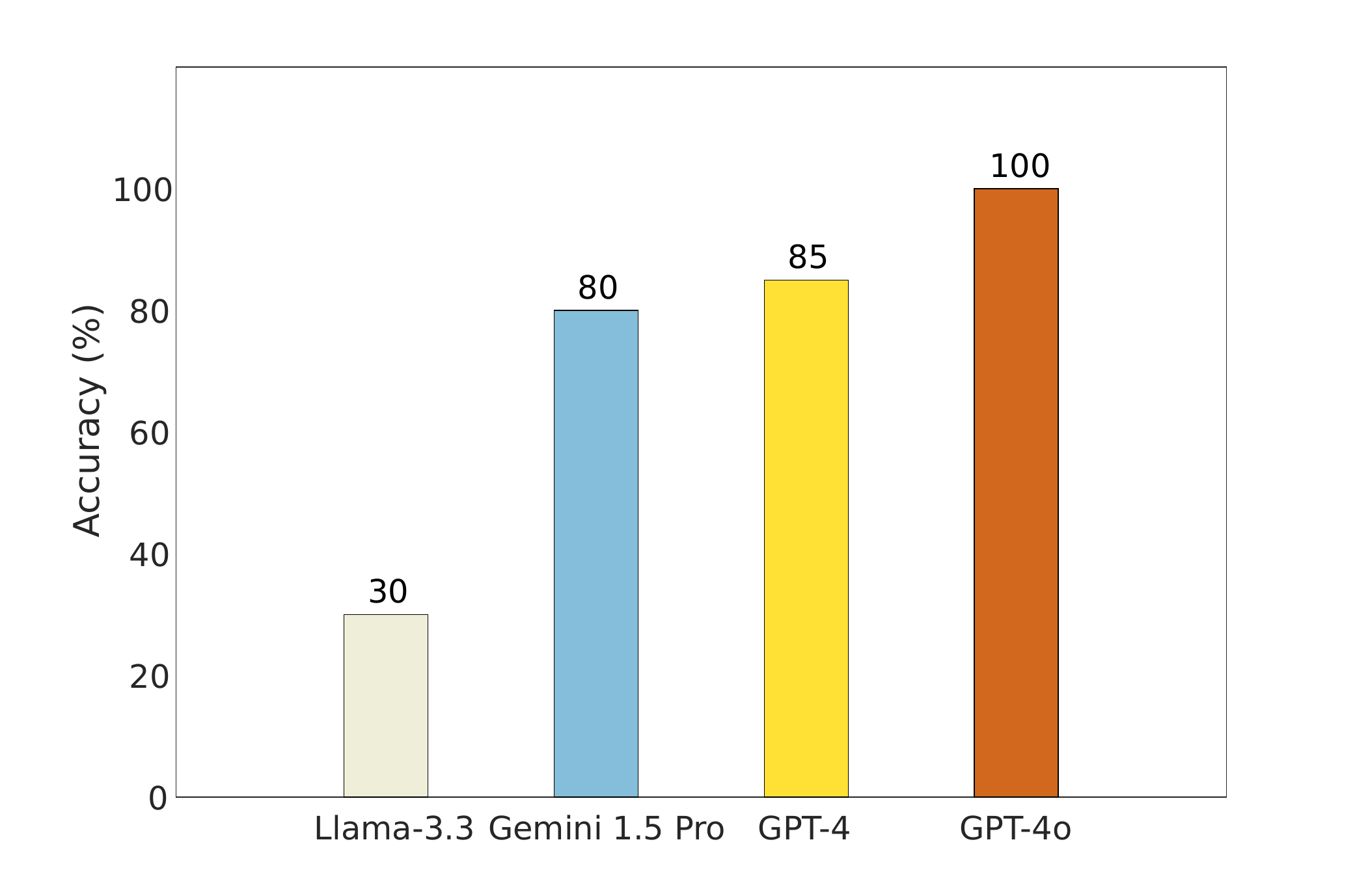}  
    \caption{Best-of-3 performance of four LLMs on the twenty manually designed examples}
    \label{multillms}
\end{figure}

Considering the inherent randomness in the generative process of LLMs, we run each experiment three consecutive times for each problem. If our framework produces a correct response in at least one of the three attempts, we consider the problem solved by the framework. We adopt a Best-of-N sampling strategy\cite{kapoor2024ai} because it better reflects the model’s potential and allows for a fair comparison of the upper-bound performance across different models. We evaluate four LLMs—Llama-3.3, Gemini 1.5 Pro, GPT-4, and GPT-4o—on 20 SAWPs. Our findings indicate that GPT-4, GPT-4o, and Gemini 1.5 Pro exhibit strong capabilities in generating Python code for structural analysis, achieving overall accuracy rates of 85\%, 100\%, and 80\%, respectively. Our framework based on these models effectively extract information from natural language problem descriptions, construct finite element models in Python, and visualize the results. The results suggest that with advancements in state-of-the-art LLMs, our framework holds great potential to assist structural engineers in structural analysis, ranging from simple to complex structures. Notably, GPT-4o, as an enhanced version of GPT-4, demonstrates significant performance improvements. This progression highlights the increasing ability of cutting-edge LLMs to automate structural analysis of simple 2D frames. With rapid advancements in LLM development, performance optimization techniques, and ongoing research into their applications in structural engineering, the role of LLMs in this field is expected to expand further, offering deeper integration and enhanced capabilities. Additionally, techniques such as few-shot learning and in-context learning can enhance the performance of LLMs in structural analysis. Among the evaluated models, GPT-4o exhibited the strongest capability in information extraction, code generation, system instruction comprehension, and space reasoning, achieving an accuracy of 100\% without any fine-tuning. As seen in Figure \ref{multillms}, our framework based on GPT-4o consistently outperformed the other models in solving the given problems within limited attempts.

We also observe that our framework demonstrates a remarkable ability to learn and generalize structural patterns from minimal examples. Building on the strong performance shown in Figure \ref{multillms}, the framework not only solved problems with high accuracy but also exhibited pattern abstraction capabilities from a single reference case. Specifically, after being provided with only one example describing the modeling of a one-bay, one-story frame (Example 1), the framework successfully extended this structural pattern to more complex configurations, such as a one-bay, two-story frame (Example 5), a two-bay, one-story frame (Example 7), and a three-bay, two-story frame (Example 9). This demonstrates the model's ability to extract high-level structural concepts—such as “bay” and “story”—from natural language and apply them accurately in code generation for structural modeling. Furthermore, the framework exhibited strong extrapolation capabilities beyond the initial instruction. While the reference example only illustrated the modeling of columns and girders, the model autonomously incorporated additional structural components, including diagonal members (Example 15 and Example 17) and cantilever beams (Example 19). These results suggest that LLMs, when guided by a structured prompting framework, are capable of reasoning from a limited seed input to produce correct and meaningful generalizations, even when faced with novel structural patterns absent from the original example.

\begin{figure}[h]
    \centering
    \includegraphics[width=1\textwidth]{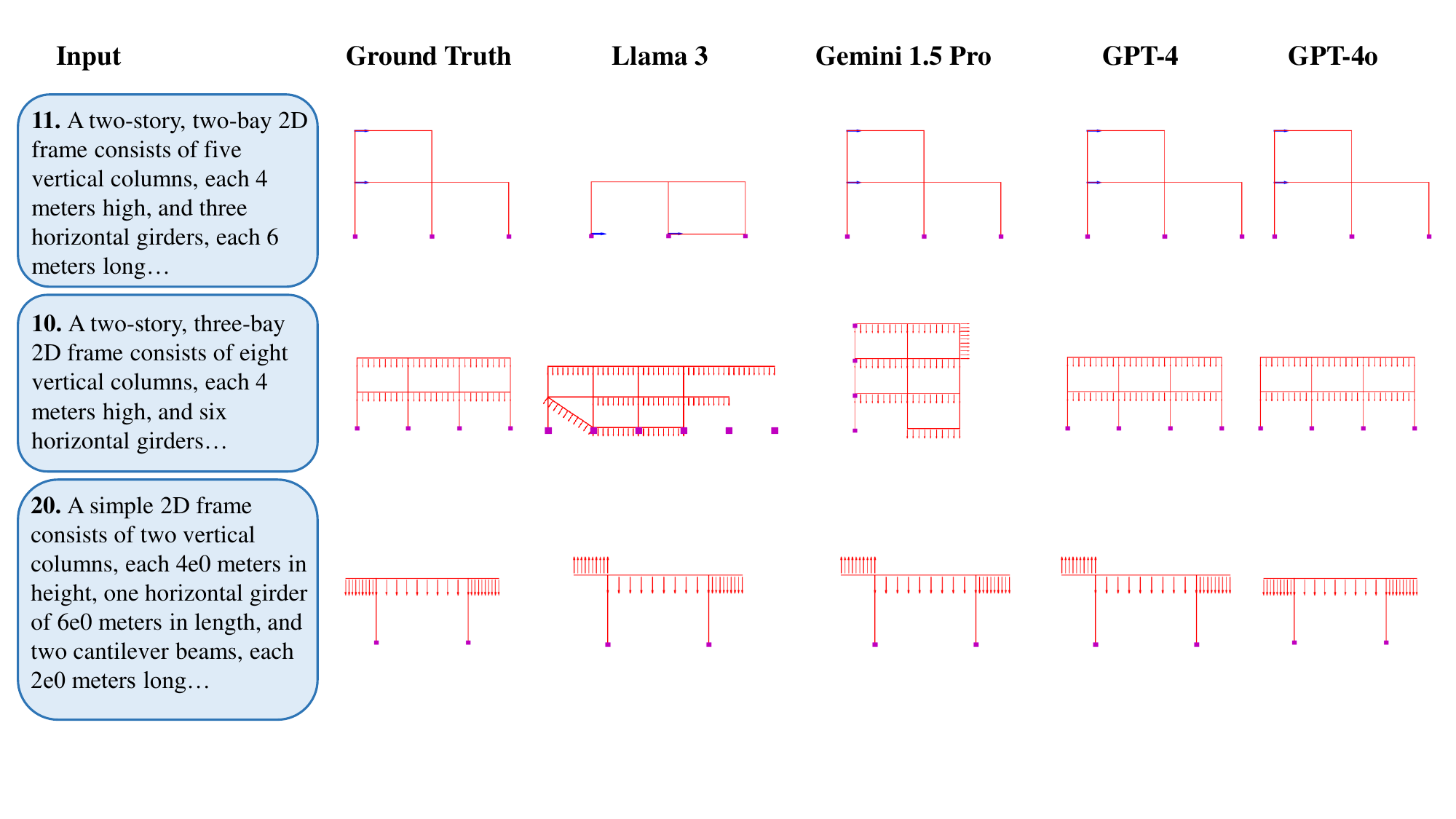}  
    \caption{Performance of four LLMs on 20 manually designed examples (detailed problem descriptions are provided in \ref{dataset}).}
    \label{multillms_visual}
\end{figure}

\FloatBarrier
As shown in Figure \ref{multillms_visual}, our framework based on different LLMs exhibit distinct strengths and weaknesses. Among the four models, llama 3 fails all three problems, highlighting its relatively weak general capability in solving these tasks. Additionally, compared to GPT-4 and GPT-4o, Gemini 1.5 Pro struggles with scaling simple structural patterns. For instance, in Example 10, when transitioning from a one-bay, one-story frame to a three-bay, two-story frame, it consistently either misdefines the structural layout or attempts to use loop statements to define nodes and elements, but the generated code is never executable. Furthermore, GPT-4o demonstrates a better understanding of system instructions than GPT-4. For example, in Example 20, when both models are given specific instruction to determine the correct direction of distributed loads, GPT-4o successfully improves its performance on this subtask, whereas GPT-4 does not.

\FloatBarrier
\subsection{Evaluation of Output Stability in the Proposed Framework}

We analyze all 20 SAWPs from the dataset in \ref{dataset} to assess the generative stability of our framework based on GPT-4o. Although these 2D frame SAWPs are relatively simple for structural engineering professionals, language models often struggle with them due to the lack of systematically curated datasets in this field during training stages and inherent limitations in space reasoning \cite{yang2024thinking}. LLMs are proficient in writing code, solving mathematical problems through reasoning, and retrieving information via web searches. However, as demonstrated in \ref{baseline}, LLMs consistently fail to solve SAWPs using standard prompting methods\cite{wei2022chain}. Standard prompting refers to directly asking the model to solve the problem without providing any additional guidance, such as solution templates or task-specific instructions. Despite the inclusion of structural analysis knowledge within their training data, LLMs struggle to effectively utilize this knowledge to accurately solve SAWPs. \ref{baseline} also provides the ground truth solutions for comparison. Notably, instruction tuning has significantly improved GPT-4o's ability to handle SAWPs. The proposed framework exhibits robust generative stability across most cases. Furthermore, in particularly challenging scenarios, the framework successfully solves the problems within a limited number of iterative attempts.

% Notably, instruction tuning significantly enhances GPT-4o's problem-solving capabilities on these tasks. The proposed framework demonstrates remarkable stability across most examples, and in some highly challenging cases, it successfully resolves the problems within a limited number of attempts.

\begin{figure}[h]
    \centering
    \includegraphics[width=1\textwidth]{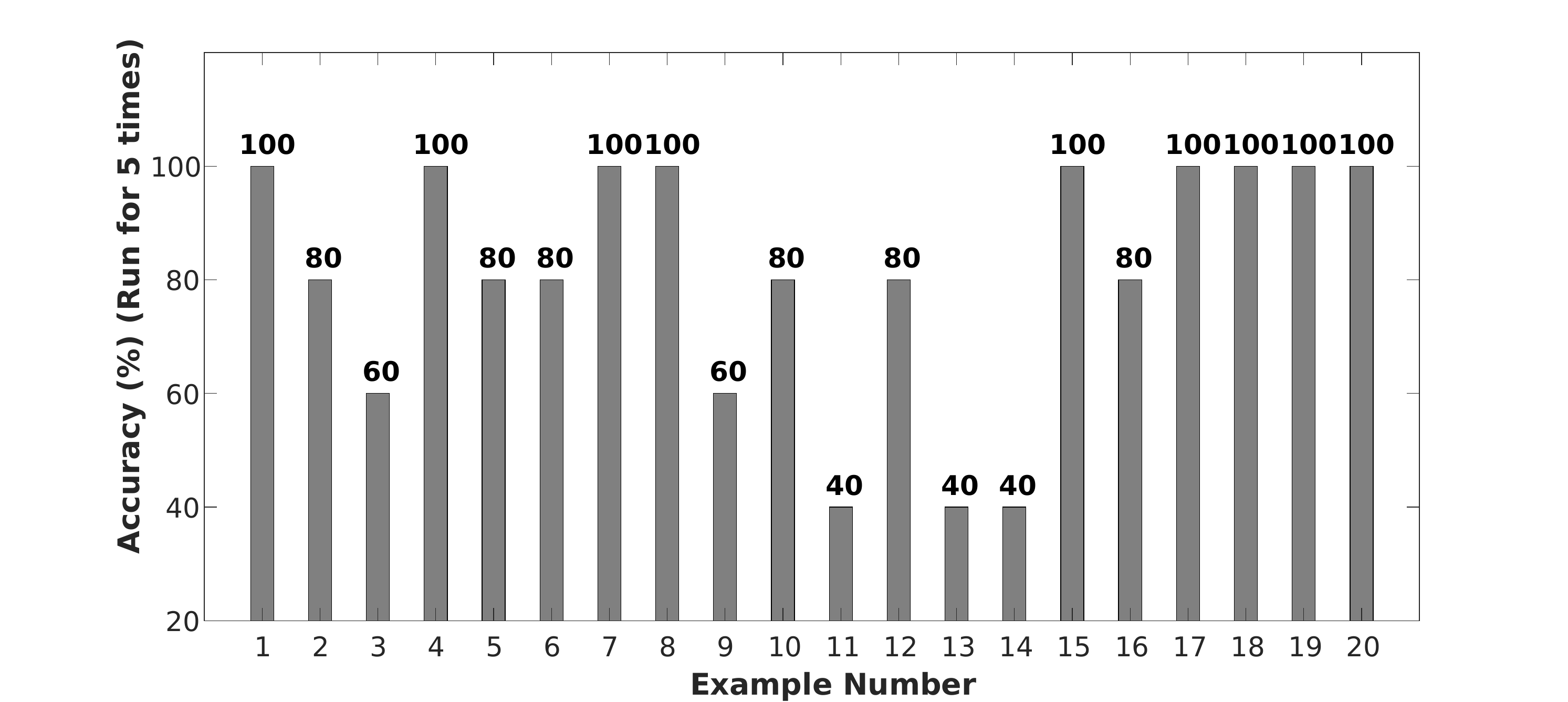}  
    \caption{Stability experiment for GPT-4o (accuracy (\%) of 20 SAWPs)}
    \label{stability}
\end{figure}

We conducted the experiment using GPT-4o within our framework, running it five times for each problem in the benchmark. The reported accuracy represents the probability that our framework successfully generates a correct and complete solution within these five attempts. The key findings from the stability experiment are summarized in Figure \ref{stability}. First, with only one reference example and limited attempts, our framework successfully generates complete and executable Python code using OpenSeesPy to perform structural analysis for all 20 SAWPs. 

However, the experiment also reveals that our framework’s performance declines when handling asymmetrical frames, as seen in Examples 11 and 13. This indicates that while LLMs can effectively learn and generalize structural patterns, they may struggle with asymmetric configurations. The detailed experimental results are further summarized in Table \ref{stability_table}.

\begin{table}[h]
    \centering
    \caption{Summary of stability experiment}
    \label{stability_table}
    \begin{tabular}{l c l}
        \toprule
        \textbf{Accuracy} & \textbf{Example Number} & \textbf{Total} \\
        \midrule
        40\%        & 11, 13, 14           & 3 \\
        60\%-80\%   & 2, 3, 5, 6, 9, 10, 12, 16 & 8 \\
        \textbf{100\%}       & 1, 4, 7, 8, 15, 17, 18, 19, 20 & 9 \\
        \bottomrule
    \end{tabular}
\end{table}

\begin{figure}[h]
    \centering
    \includegraphics[width=1\textwidth]{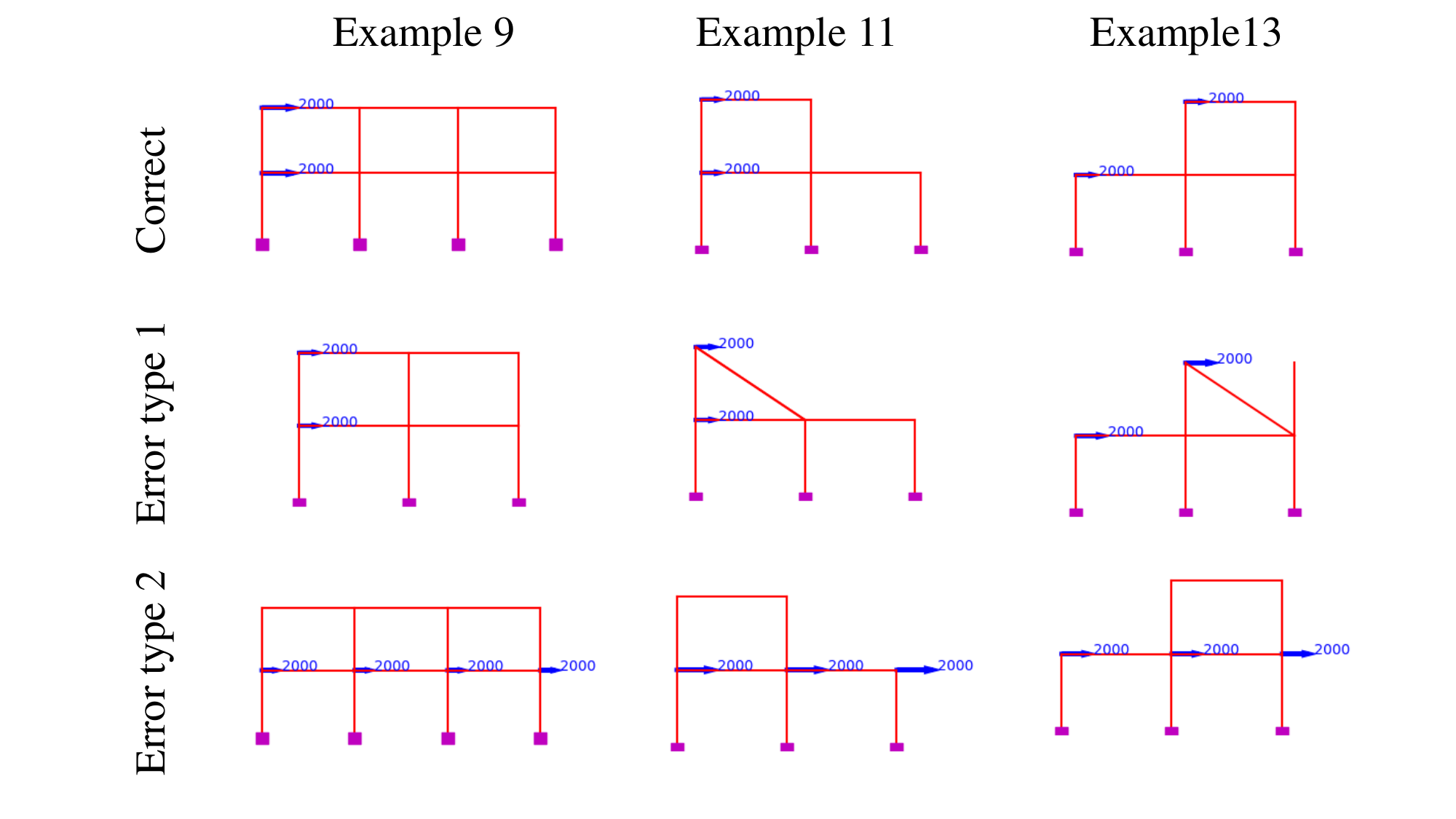}  
    \caption{GPT-4o utilizes OpenSeesPy and OpsVis to generate results based on user input}
    \label{openseespy}
\end{figure}

As illustrated in Figure \ref{openseespy}, when GPT-4o generates runnable code, some scripts fail due to coding coherence issues and other inconsistencies, primarily making two types of mistakes: Error Type 1, where the framework fails to sketch the structural layout as required. For instance, in Example 9, the frame should be a three-bay, two-story frame, but GPT-4o generates a two-bay, two-story frame; in Example 11, it fails to define a required node, resulting in an incomplete second-story frame; and in Example 13, it incorrectly defines an element, leading to an inaccurate second story. Error Type 2 involves incorrect definitions of boundary conditions, where the framework misassigns loads. In the three given examples, GPT-4o was instructed to assign point loads to the left side of the frame but instead assigned them to all nodes on the first floor; in other cases, it may also misdefine structural supports. We identify two main causes of these errors: first, at this stage, both input and output are limited to text (including code), meaning GPT-4o cannot "see" the results it generates, preventing it from recognizing misassigned elements or loads. In future work, we may incorporate multimodal capabilities into our framework by integrating a validation layer that uploads graphical results during the initial generation phase—if the LLM detects an error pattern, it can provide modification suggestions to update the generated output. Second, we currently teach LLMs only the correct structural patterns without exposing them to incorrect ones. Due to the inherent randomness of LLM-generated outputs, a single problem may yield multiple structural layouts, some correct and others incorrect. To mitigate this, future work can incorporate negative sampling techniques to reduce the likelihood of generating incorrect structural patterns when modeling frames from natural language input.

\FloatBarrier
\subsection{Impact of Instructions on LLM Output Quality}

Although our framework can understand SAWPs descriptions and generate Python scripts through referencing only a simple example and its corresponding code, it still makes many basic mistakes due to its lack of space reasoning capabilities for structures in 2D space. To address this, we provide structured system instructions to help the language model align with human understanding of structural layouts in space. The complete set of system instructions can be found in Table \ref{full_instructions}. To be more specific, Direction reasoning provides instructions on how to judge the direction of the structure. Humans have a unique visual system, allowing them to determine the orientation of a structure without explicit reasoning. However, the LLMs we are using lack visual recognition capabilities. Instead, they have access to the coordinates of all structural nodes. This component instructs LLMs on how to judge the direction of the structure using node coordinates to accurately define loads or supports. Furthermore, we also introduce number reasoning in the system instructions. Before adding this instruction, we found that LLMs often defined elements with a different number than what was mentioned in the problem description. To address this issue, this instruction ensures that LLMs define the correct number of elements in the structures.

\begin{table}[h]
    \centering
    \renewcommand{\arraystretch}{1.3}
    \caption{Structural reasoning instructions}
    \label{full_instructions}
    \begin{tabular}{p{4cm} p{11cm}}
        \toprule
        \textbf{Category} & \textbf{Description} \\
        \midrule
        \textbf{Direction reasoning} & 
        To correctly apply loads to specific regions of the structure, follow these steps: 
        \begin{enumerate}
            \item Identify all nodes in the structure, excluding support nodes. 
            \item Determine the location where the load should be applied based on coordinate values: 
            \begin{itemize}
                \item Left side: Nodes with the smallest x-coordinate.  
                \item Right side: Nodes with the largest x-coordinate.  
                \item Top side: Nodes with the largest y-coordinate.  
                \item Bottom side: Nodes with the smallest y-coordinate.  
            \end{itemize}
            \item Assign the load to the identified nodes accordingly.  
        \end{enumerate} \\
        \midrule
        \textbf{Number reasoning} & 
        To ensure the structure contains the correct number of members:  
        \begin{enumerate}
            \item Identify all vertical members (columns) in the structure.  
            \item Verify that the number of defined columns matches the stated count in the problem.  
            \item Similarly, count all horizontal members (girders) and ensure their number aligns with the problem description.  
        \end{enumerate} \\
        \midrule
        \textbf{Space rationality reasoning} & 
        To maintain spatial consistency in structural elements:  
        \begin{enumerate}
            \item By default, set the y-coordinate of support nodes to zero unless specified otherwise.  
            \item Ensure horizontal members (girders) have two nodes with identical y-coordinates.  
            \item Ensure vertical members (columns) have two nodes with identical x-coordinates.  
            \item If diagonal members are not mentioned, ensure that no element has nodes with both x and y coordinates differing.  
        \end{enumerate} \\
        \midrule
        \textbf{Distributed loading direction reasoning} & 
        When applying a distributed load to an element:  
        \begin{enumerate}
            \item Check the direction of the load: If the load is inward, apply a negative sign to the load value.  
            \item Identify the starting and ending nodes of the element.  
            \item If the x-coordinate of the starting node is smaller than the ending node, assign the load as given.  
            \item If the x-coordinate of the starting node is greater than the ending node, negate the distributed load value before applying it.  
        \end{enumerate} \\
        \bottomrule
    \end{tabular}
\end{table}

\begin{figure}[h]
    \centering
    % Place the first and third images on the left side (stacked vertically)
    \begin{minipage}{0.35\textwidth} % Adjust width as needed
        \centering
        \includegraphics[width=\textwidth]{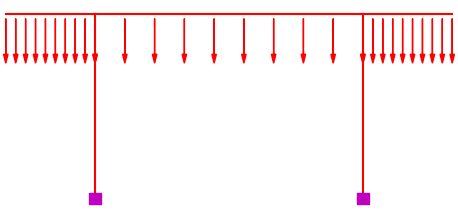}
        \caption{Positive result \label{reasoning_1_1}}
        
        \vspace{0.35cm} % Spacing between images
        \includegraphics[width=\textwidth]{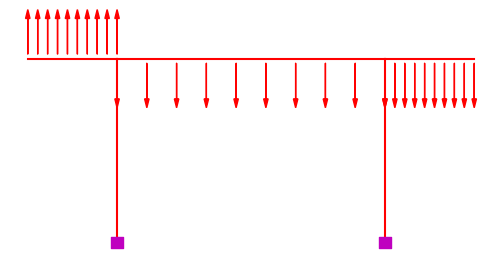}
        \caption{Negative result \label{reasoning_1_2}}
    \end{minipage}%
    \hspace{0.02\textwidth} % Horizontal spacing
    % Place the second image on the right side
    \begin{minipage}{0.6\textwidth} % Adjust width as needed
        \centering
        \includegraphics[width=\textwidth,trim=0 0 0 170, clip]{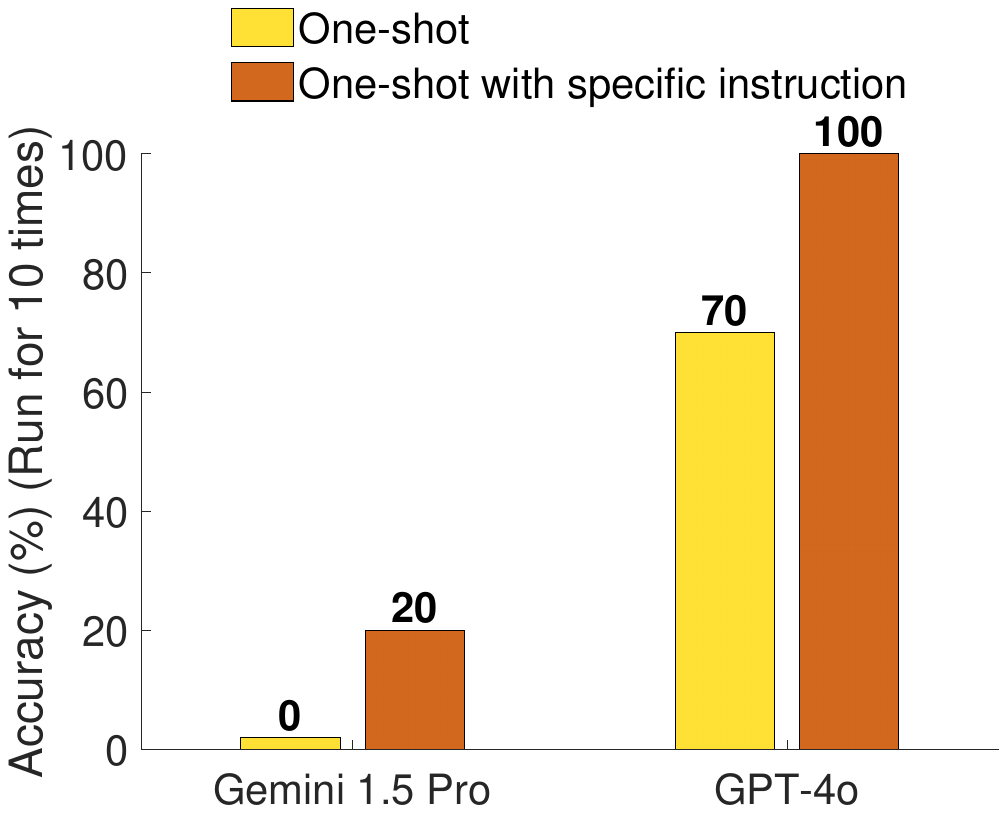}
        \vspace{-4cm}
        \caption{LLMs, such as Gemini 1.5 Pro and GPT-4o, are optimized with distributed loading direction reasoning instruction to enhance their accuracy in applying distributed loads to the structure in problem 20. \label{resoning_comp_1}}
    \end{minipage}
\end{figure}

In addition, in problem number 20 from Table \ref{schematic_partial}, the correct ground truth should be as shown in Figure \ref{reasoning_1_1}. When defining the distributed load on an element, engineers must be particularly cautious about the load direction, as it is not always the same. The sign preceding the load value depends on how the element is defined. Therefore, ensuring consistency between the element definition and load definition is crucial for accurate modeling. In this problem, before we explicitly provide instructions on how to determine the correct sign of the load value, the framework is likely to make an error in defining the direction of the distributed load, as illustrated in Figure \ref{reasoning_1_2}. To prevent such mistakes, we include a specific instruction named distributed loading direction reasoning in Table \ref{full_instructions}, explaining how structural engineers approach this problem. As seen in Figure \ref{resoning_comp_1}, we provided the problem description corresponding to problem number 20 to both Gemini 1.5 Pro and GPT-4o for ten times independently. We found that Gemini 1.5 Pro and GPT-4o demonstrate improved performance in correctly applying distributed loads on structures when guided by the specific instruction. Without this instruction, Gemini failed to define the distributed load correctly, while GPT-4o achieved an accuracy of 70\%. However, after incorporating the specific instruction, our framework based on both models showed performance improvements on this task, with GPT-4o reaching 100\% accuracy. The results demonstrate that correct and precise instructions from experienced human experts can enhance LLMs' understanding and performance in solving structural analysis problems. However, the underlying mechanisms of this improvement require further interpretability research in this domain. Additionally, systematic ablation experiments with more instructions on larger datasets should be conducted to accurately assess the impact of specific instructions on overall model performance.

\begin{figure}[h]
    \centering
    \begin{subfigure}{0.63\textwidth}
        \centering
        \includegraphics[width=\textwidth,trim=80 0 80 0, clip]{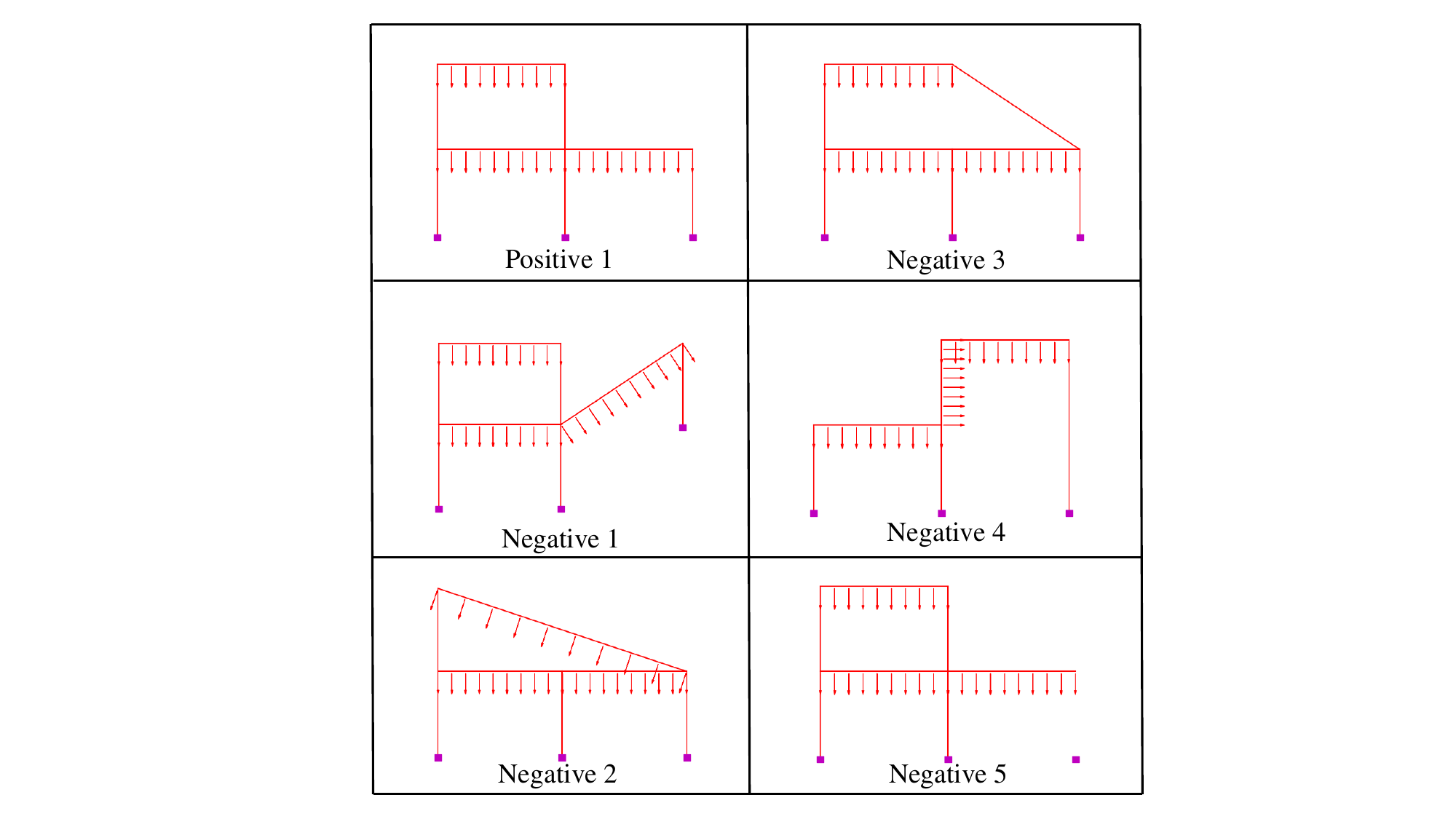}
        \caption{Cluster generated by GPT-4o for the solution of example 12}
        \label{cluster_12}
    \end{subfigure}
    \hfill
    \begin{subfigure}{0.35\textwidth}
        \centering
        \includegraphics[width=\textwidth,trim=100 170 100 120, clip]{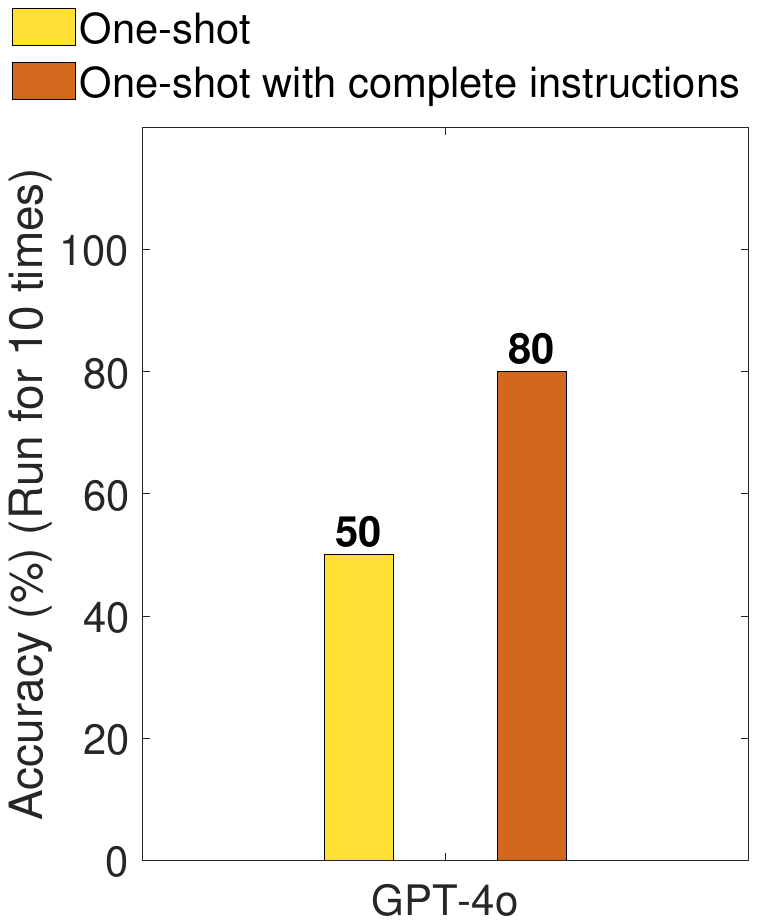}
        \caption{Impact of complete instructions}
        \label{instructiontuning}
    \end{subfigure}
    \caption{Comparison of clustering results and instruction tuning}
    \label{fig:comparison}
\end{figure}

For a more complex example like Example 12 in Figure \ref{schematic_partial}, if instructions in Table \ref{full_instructions} are not provided for the LLMs, the framework generates a cluster of results, as shown in Figure \ref{cluster_12}. These results contain various types of errors, such as incorrect layout definitions, incorrect load applications, incorrect support definitions, or a combination of these mistakes. The primary reason for these errors is the increased number of nodes and elements, along with the asymmetry of the structure. To address this, we incorporated the complete instructions from Table \ref{full_instructions} into the system instructions to evaluate whether they could enhance the framework’s performance on this task. As shown in Figure \ref{instructiontuning}, we provided the problem description of Example 12 to GPT-4o and run the experiment ten times, with and without the complete system instructions. We found that including the instructions increases the framework's accuracy on this problem from 50\% to 80\%. However, continuously expanding system instructions is not always beneficial. Increasing the number of instructions introduces a scalability dilemma: more instructions lead to higher token consumption, increasing computational costs and resource demands for maintaining an efficient generative process. Additionally, excessive instructions can confuse LLMs, as they lack weighted prioritization to distinguish more critical directives from less relevant ones. Ambiguous instructions may even degrade overall performance compared to experiments conducted without any system instructions. Therefore, determining optimal instruction combinations and developing efficient methods for synthesizing and compressing system instructions could be valuable directions for future research.

\FloatBarrier
\section{Concluding Remarks}\label{conclusion}

We explored the potential of LLMs to perform structural analysis based solely on natural language input. We developed a framework using structured prompt design and in-context learning strategies to better align LLMs with the specific needs of structural engineering. By integrating LLMs with the finite element analysis tool OpenSeesPy, the framework reduces the complexity of using domain-specific software and minimizes repetitive tasks involved in creating or modifying models. We also manually designed 20 structural analysis word problems incorporating ground truth, problem descriptions, and corresponding code to serve as a benchmark for evaluating different LLMs’ problem-solving capabilities. Using this benchmark, we conducted experiments on comparative analysis of different LLMs, model stability, and enhancement techniques, including the impact of additional system instructions on reasoning and coding ability. Major conclusions are as follows:

\begin{enumerate}
    \item GPT-4o demonstrates the best performance among the evaluated models on our benchmark, showing strong capability in solving domain-specific structural analysis problems based on natural language input.
    \item Reasoning instructions significantly enhance performance, especially for tasks involving code generation and reasoning in structural engineering contexts, indicating the importance of prompt design in aligning LLM behavior with engineering goals.
    \item LLMs demonstrate strong potential to become an integral component of structural engineering workflows by enhancing efficiency, reducing costs, and streamlining repetitive modeling tasks. Their capacity to generate accurate structural models directly from natural language input further enables the development of LLM-based systems that support rapid and reliable decision-making in real-world, time-sensitive engineering applications.
\end{enumerate}

While our results demonstrate the potential of LLM-based structural analysis, several important limitations must be addressed before widespread adoption in practice. First, our dataset size and the number of experimental runs were limited since the dataset was manually designed for this project while also considering the usage of APIs. A key question for future research is how to obtain larger datasets and invoke LLMs' APIs at lower costs in this domain. Second, although the cost of manually augmenting exemplars with system instructions is minimal in the few-shot setting, such annotation costs could be prohibitive for fine-tuning, and the cost of expert-level evaluation of LLM-generated results could be very high (though this could potentially be surmounted with synthetic data generation~\cite{zhang2022automatic} and LLMs as judges~\cite{li2024llms}). Third, there is no guarantee that the system instructions are always correct, which can lead to both accurate and inaccurate responses. For instance, after multiple runs on a single task, we observed that the model continued to make similar mistakes—despite the system instructions explicitly addressing solutions to avoid these potential errors. Improving the alignment of tactical approaches with structural engineering needs remains a promising direction for future research. Finally, this work only investigates LLMs' performance on manually designed tasks; further research could explore how to effectively integrate LLMs into real-world applications in this sector. Future work could involve incorporating advanced techniques such as supervised fine-tuning (SFT) and reinforcement learning fine-tuning (RLFT) to further enhance LLMs' reasoning and tool-using capabilities for structural analysis. In addition, integrating LLMs specialized in structural analysis with LLMs which are proficient in structural design has the potential to create agentic structural engineers for real-world applications.

\clearpage
\bibliographystyle{unsrt}
\bibliography{references}

\clearpage
\appendix
\renewcommand{\thesection}{Appendix \Alph{section}}  
\section{Dataset} \label{dataset}

\begin{table}[h]
    \centering
    \renewcommand{\arraystretch}{1.5}
    \caption{Problem Descriptions and Ground Truth Schematics}
    \label{schematic_1}
    \setlength{\extrarowheight}{2pt}
    \renewcommand{\tabularxcolumn}[1]{>{\hsize=\dimexpr\hsize+2\tabcolsep\relax\justifying\arraybackslash}m{#1}}
    \begin{tabularx}{\textwidth}{|X|>{\centering\arraybackslash}m{0.32\textwidth}|} 
        \hline
        \textbf{Problem description} & \textbf{Ground truth} \\
        \hline
        \textbf{1. }How does a simple 2D frame, consisting of two vertical columns (\(4 \times 10^0\) meters in height) and one horizontal girder (\(6 \times 10^0\) meters in length), behave under the combined effects of a horizontal point load of \(2 \times 10^3\) N at the top of one column and a uniform vertical distributed load of \(1 \times 10^4\) N/m along the girder, considering elastic material properties with a Young's modulus \(E = 2 \times 10^{11}\) Pa, column cross-sectional area \(A_c = 2 \times 10^{-3} \, \text{m}^2\), girder cross-sectional area \(A_g = 6 \times 10^{-3} \, \text{m}^2\), column moment of inertia \(I_c = 1.6 \times 10^{-5} \, \text{m}^4\), and girder moment of inertia \(I_g = 5.4 \times 10^{-5} \, \text{m}^4\). All supports are fixed. What are the resulting deformations and internal forces (axial, shear, and moment) in the frame?
        & \begin{minipage}[t][4cm][t]{0.32\textwidth}
            \centering
            \includegraphics[scale=0.5]{images/schematic_01.png}
        \end{minipage} \\
        \hline
        \textbf{2. }How does a simple 2D frame, consisting of one vertical column (\(4 \times 10^0\) meters in height), one diagonal member forming the brace in the left side of the column, where the horizontal height from the top of the column to the support of the diagonal member is \(6 \times 10^0\) meters, behave under the effect of a horizontal point load of \(2 \times 10^3\) N at the top of the column, considering elastic material properties with a Young's modulus \(E = 2 \times 10^{11}\) Pa, column cross-sectional area of \(2 \times 10^{-3}\) m\(^2\), diagonal member cross-sectional area of \(6 \times 10^{-3}\) m\(^2\), column moment of inertia of \(1.6 \times 10^{-5}\) m\(^4\), and diagonal member moment of inertia of \(5.4 \times 10^{-5}\) m\(^4\) and all supports are fixed. What are the resulting deformations and internal forces (axial, shear, and moment) in the frame?
        & \begin{minipage}[t][4cm][t]{0.32\textwidth}
            \centering
            \includegraphics[scale=0.5]{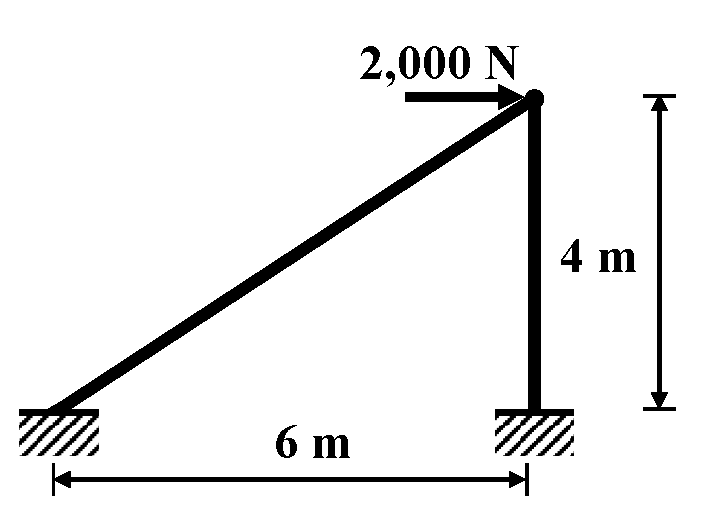}
        \end{minipage} \\
        \hline
        \textbf{3. }How does a simple 2D frame, consisting of two vertical columns (\(4 \times 10^0\) meters in height) and one horizontal girder (\(6 \times 10^0\) meters in length), behave under the effect of a horizontal point load of \(2 \times 10^3\) N at the top of one column, considering elastic material properties with a Young's modulus \(E = 2 \times 10^{11}\) Pa, column cross-sectional area of \(2 \times 10^{-3}\) m\(^2\), girder cross-sectional area of \(6 \times 10^{-3}\) m\(^2\), column moment of inertia \(I_c = 1.6 \times 10^{-5}\) m\(^4\), and girder moment of inertia \(I_g = 5.4 \times 10^{-5}\) m\(^4\) and all supports are fixed. What are the resulting deformations and internal forces (axial, shear, and moment) in the frame?
        & \begin{minipage}[t][4cm][t]{0.32\textwidth}
            \centering
            \includegraphics[scale=0.5]{images/schematic_03.png}
        \end{minipage} \\
        \hline
        \textbf{4. }How does a simple 2D frame, consisting of two vertical columns (\(4 \times 10^0\) meters in height) and one horizontal girder (\(6 \times 10^0\) meters in length), behave under the effect of a uniform vertical distributed load of \(1 \times 10^4\) N/m along the girder, considering elastic material properties with a Young's modulus \(E = 2 \times 10^{11}\) Pa, column cross-sectional area of \(2 \times 10^{-3}\) m\(^2\), girder cross-sectional area of \(6 \times 10^{-3}\) m\(^2\), column moment of inertia \(I_c = 1.6 \times 10^{-5}\) m\(^4\), and girder moment of inertia \(I_g = 5.4 \times 10^{-5}\) m\(^4\) and all supports are fixed. What are the resulting deformations and internal forces (axial, shear, and moment) in the frame?
        & \begin{minipage}[t][4cm][t]{0.32\textwidth}
            \centering
            \includegraphics[scale=0.5]{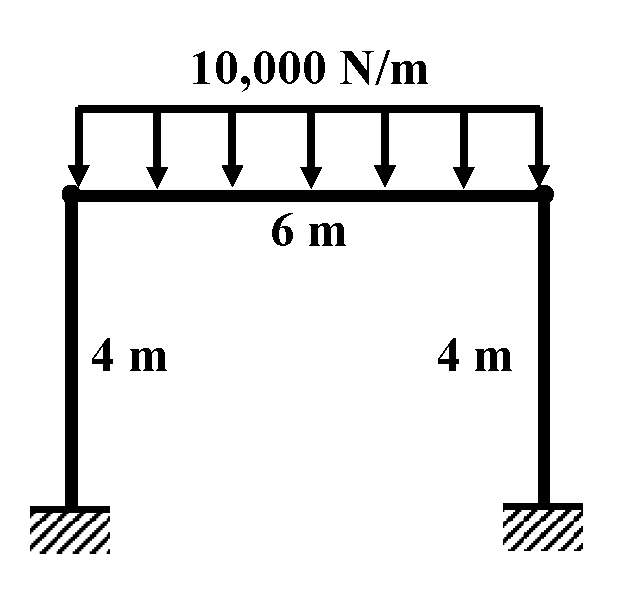}
        \end{minipage} \\
        \hline
    \end{tabularx}
\end{table}

\begin{table}[h]
    \centering
    \renewcommand{\arraystretch}{1.5}
    \caption{Additional Problem Descriptions and Ground Truth Schematics}
    \label{schematic_2}
    \setlength{\extrarowheight}{2pt}
    \renewcommand{\tabularxcolumn}[1]{>{\hsize=\dimexpr\hsize+2\tabcolsep\relax\justifying\arraybackslash}m{#1}}
    \begin{tabularx}{\textwidth}{|X|>{\centering\arraybackslash}m{0.32\textwidth}|} 
        \hline
        \textbf{Problem description} & \textbf{Ground truth} \\
        \hline
        \textbf{5. }How does a two-story 2D frame, consisting of four vertical columns (\(4\) meters in height for each story) and two horizontal girders (\(6\) meters in length each), behave under the horizontal point load of \(2 \times 10^3\) N at each column on the left side? Consider elastic material properties with Young's modulus of \(2 \times 10^{11}\) Pa, column cross-sectional area of \(2 \times 10^{-3}\) m\(^2\), girder cross-sectional area of \(6 \times 10^{-3}\) m\(^2\), column moment of inertia of \(1.6 \times 10^{-5}\) m\(^4\), and girder moment of inertia of \(5.4 \times 10^{-5}\) m\(^4\). All supports are fixed. What are the resulting deformations and internal forces (axial, shear, and bending moment) within the frame?
        & \begin{minipage}[t][4cm][t]{0.32\textwidth}
            \centering
            \includegraphics[scale=0.4]{images/schematic_05.png}
        \end{minipage} \\
        \hline
        \textbf{6. }How does a two-story 2D frame, consisting of four vertical columns (\(4\) meters in height for each story) and two horizontal girders (\(6\) meters in length each), behave under the uniform vertical distributed load of \(1 \times 10^4\) N/m along each girder? Consider elastic material properties with Young's modulus of \(2 \times 10^{11}\) Pa, column cross-sectional area of \(2 \times 10^{-3}\) m\(^2\), girder cross-sectional area of \(6 \times 10^{-3}\) m\(^2\), column moment of inertia of \(1.6 \times 10^{-5}\) m\(^4\), and girder moment of inertia of \(5.4 \times 10^{-5}\) m\(^4\). All supports are fixed. What are the resulting deformations and internal forces (axial, shear, and bending moment) within the frame?
        & \begin{minipage}[t][4cm][t]{0.32\textwidth}
            \centering
            \includegraphics[scale=0.4]{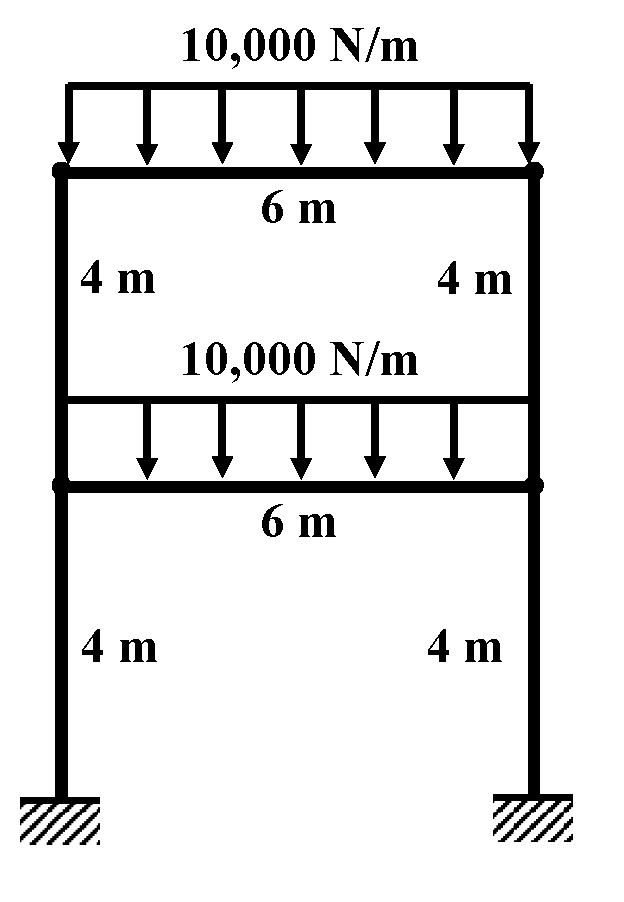}
        \end{minipage} \\
        \hline
        \textbf{7. }How does a one-story two-bay 2D frame, consisting of three vertical columns (\(4\) meters in height each) and two horizontal girders (\(6\) meters in length each), behave under the horizontal point load of \(2 \times 10^3\) N at the top of the column on the left side? Consider elastic material properties with Young's modulus of \(2 \times 10^{11}\) Pa, column cross-sectional area of \(2 \times 10^{-3}\) m\(^2\), girder cross-sectional area of \(6 \times 10^{-3}\) m\(^2\), column moment of inertia of \(1.6 \times 10^{-5}\) m\(^4\), and girder moment of inertia of \(5.4 \times 10^{-5}\) m\(^4\). All supports are fixed. What are the resulting deformations and internal forces (axial, shear, and bending moment) within the frame?
        & \begin{minipage}[t][4cm][t]{0.32\textwidth}
            \centering
            \includegraphics[scale=0.35]{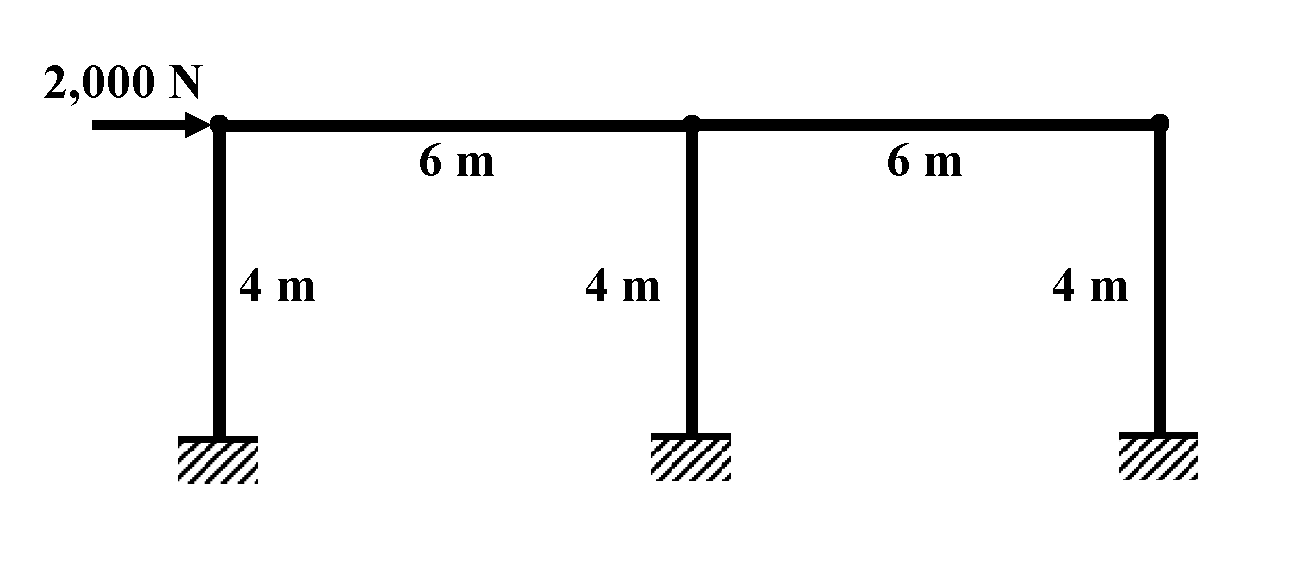}
        \end{minipage} \\
        \hline
        \textbf{8. }How does a one-story two-bay simple 2D frame, consisting of three vertical columns (\(4\) meters in height each) and two horizontal girders (\(6\) meters in length each), behave under the uniform vertical distributed load of \(1 \times 10^4\) N/m along each girder? Consider elastic material properties with Young's modulus of \(2 \times 10^{11}\) Pa, column cross-sectional area of \(2 \times 10^{-3}\) m\(^2\), girder cross-sectional area of \(6 \times 10^{-3}\) m\(^2\), column moment of inertia of \(1.6 \times 10^{-5}\) m\(^4\), and girder moment of inertia of \(5.4 \times 10^{-5}\) m\(^4\). All supports are fixed. What are the resulting deformations and internal forces (axial, shear, and bending moment) within the frame?
        & \begin{minipage}[t][4cm][t]{0.32\textwidth}
            \centering
            \includegraphics[scale=0.35]{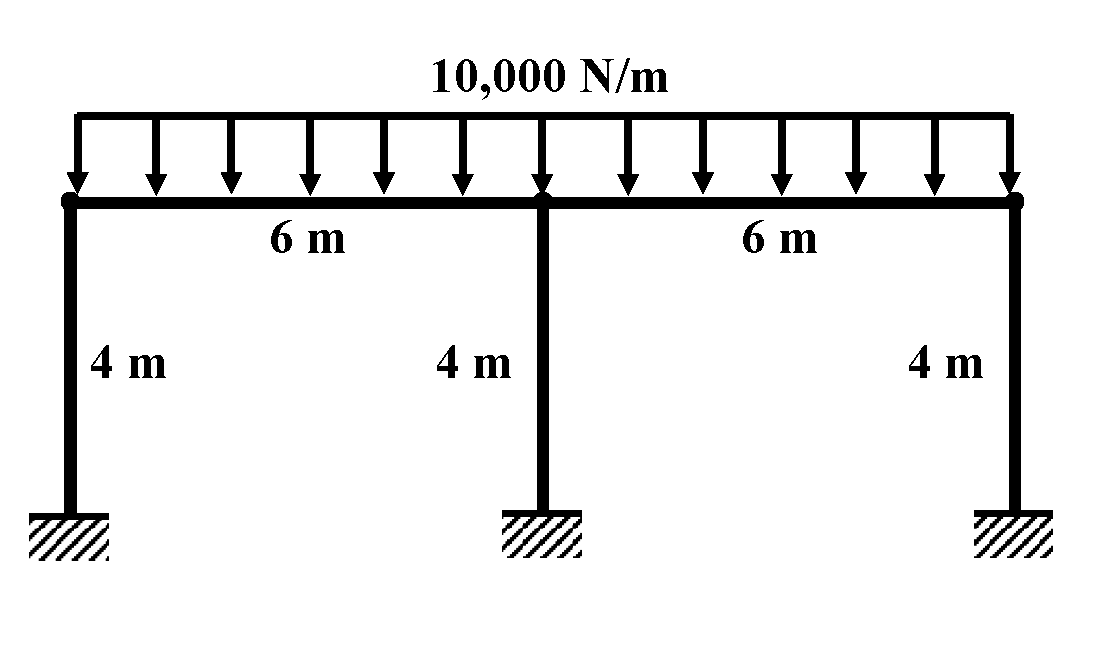}
        \end{minipage} \\
        \hline
    \end{tabularx}
\end{table}

% Third part of the table
\begin{table}[h]
    \centering
    \renewcommand{\arraystretch}{1.5}
    \caption{Additional Problem Descriptions and Ground Truth Schematics}
    \label{schematic_3}
    \setlength{\extrarowheight}{2pt}
    \renewcommand{\tabularxcolumn}[1]{>{\hsize=\dimexpr\hsize+2\tabcolsep\relax\justifying\arraybackslash}m{#1}}
    \begin{tabularx}{\textwidth}{|X|>{\centering\arraybackslash}m{0.32\textwidth}|} 
        \hline
        \textbf{Problem description} & \textbf{Ground truth} \\
        \hline
        \textbf{9. }How does a two-story three-bay 2D frame, consisting of 8 vertical columns (\(4\) meters in height each) and 6 horizontal girders (\(6\) meters in length each), behave under the horizontal point load of \(2 \times 10^3\) N at each column on the left side? Consider elastic material properties with Young's modulus of \(2 \times 10^{11}\) Pa, column cross-sectional area of \(2 \times 10^{-3}\) m\(^2\), girder cross-sectional area of \(6 \times 10^{-3}\) m\(^2\), column moment of inertia of \(1.6 \times 10^{-5}\) m\(^4\), and girder moment of inertia of \(5.4 \times 10^{-5}\) m\(^4\). All supports are fixed. What are the resulting deformations and internal forces (axial force, shear force, and bending moment) within the frame?
        & \begin{minipage}[t][4cm][t]{0.32\textwidth}
            \centering
            \includegraphics[scale=0.26]{images/schematic_09.png}
        \end{minipage} \\
        \hline
        \textbf{10. }How does a two-story three-bay 2D frame, consisting of 8 vertical columns (\(4\) meters in height each) and 6 horizontal girders (\(6\) meters in length each), behave under the uniform vertical distributed load of \(1 \times 10^4\) N/m along each girder? Consider elastic material properties with Young's modulus of \(2 \times 10^{11}\) Pa, column cross-sectional area of \(2 \times 10^{-3}\) m\(^2\), girder cross-sectional area of \(6 \times 10^{-3}\) m\(^2\), column moment of inertia of \(1.6 \times 10^{-5}\) m\(^4\), and girder moment of inertia of \(5.4 \times 10^{-5}\) m\(^4\). All supports are fixed. What are the resulting deformations and internal forces (axial force, shear force, and bending moment) within the frame?
        & \begin{minipage}[t][4cm][t]{0.32\textwidth}
            \centering
            \includegraphics[scale=0.26]{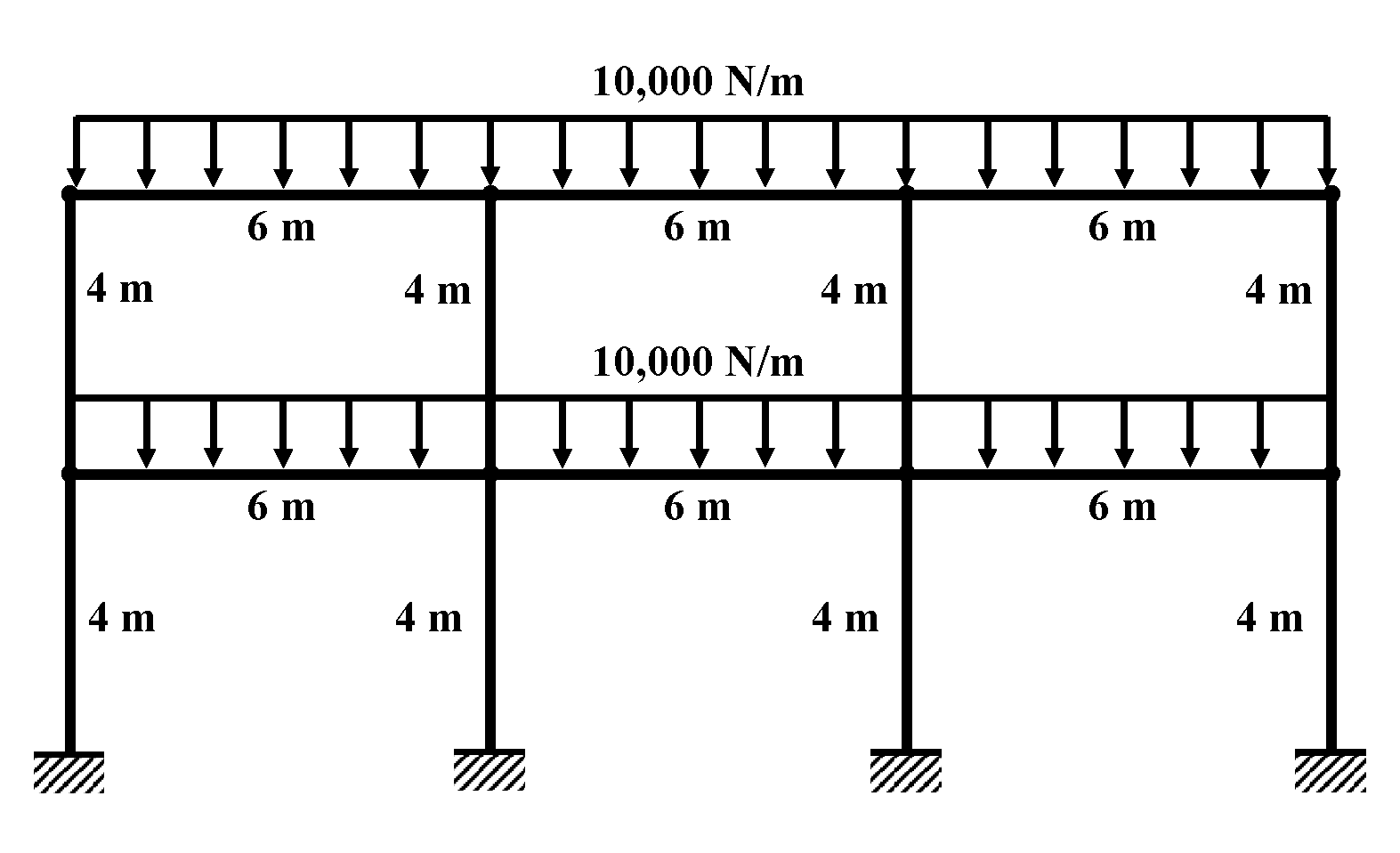}
        \end{minipage} \\
        \hline
        \textbf{11. }How does a two-story two-bay 2D frame, where the first bay has two stories and the second bay has one story, consisting of 5 vertical columns (\(4\) meters in height each) and 3 horizontal girders (\(6\) meters in length each), behave under the horizontal point load of \(2 \times 10^3\) N at each column on the left side? Consider elastic material properties with Young's modulus of \(2 \times 10^{11}\) Pa, column cross-sectional area of \(2 \times 10^{-3}\) m\(^2\), girder cross-sectional area of \(6 \times 10^{-3}\) m\(^2\), column moment of inertia of \(1.6 \times 10^{-5}\) m\(^4\), and girder moment of inertia of \(5.4 \times 10^{-5}\) m\(^4\). All supports are fixed. What are the resulting deformations and internal forces (axial force, shear force, and bending moment) within the frame?
        & \begin{minipage}[t][4cm][t]{0.32\textwidth}
            \centering
            \includegraphics[scale=0.35]{images/schematic_11.png}
        \end{minipage} \\
        \hline
        \textbf{12. }How does a two-story two-bay 2D frame, where the first bay has two stories and the second bay has one story, consisting of 5 vertical columns (\(4\) meters in height each) and 3 horizontal girders (\(6\) meters in length each), behave under the uniform vertical distributed load of \(1 \times 10^4\) N/m along each girder? Consider elastic material properties with Young's modulus of \(2 \times 10^{11}\) Pa, column cross-sectional area of \(2 \times 10^{-3}\) m\(^2\), girder cross-sectional area of \(6 \times 10^{-3}\) m\(^2\), column moment of inertia of \(1.6 \times 10^{-5}\) m\(^4\), and girder moment of inertia of \(5.4 \times 10^{-5}\) m\(^4\). All supports are fixed. What are the resulting deformations and internal forces (axial force, shear force, and bending moment) within the frame?
        & \begin{minipage}[t][4cm][t]{0.32\textwidth}
            \centering
            \includegraphics[scale=0.35]{images/schematic_12.png}
        \end{minipage} \\
        \hline
    \end{tabularx}
\end{table}

% Fourth part of the table
\begin{table}[h]
    \centering
    \renewcommand{\arraystretch}{1.5}
    \caption{Additional Problem Descriptions and Ground Truth Schematics}
    \label{schematic_4}
    \setlength{\extrarowheight}{2pt}
    \renewcommand{\tabularxcolumn}[1]{>{\hsize=\dimexpr\hsize+2\tabcolsep\relax\justifying\arraybackslash}m{#1}}
    \begin{tabularx}{\textwidth}{|X|>{\centering\arraybackslash}m{0.32\textwidth}|} 
        \hline
        \textbf{Problem description} & \textbf{Ground truth} \\
        \hline
        \textbf{13. }How does a two-story two-bay 2D frame, where the first bay has one story and the second bay has two stories, consisting of 5 vertical columns (\(4\) meters in height each) and 3 horizontal girders (\(6\) meters in length each), behave under the horizontal point load of \(2 \times 10^3\) N at the column on the left side on the first story and on the second story? Consider elastic material properties with Young's modulus of \(2 \times 10^{11}\) Pa, column cross-sectional area of \(2 \times 10^{-3}\) m\(^2\), girder cross-sectional area of \(6 \times 10^{-3}\) m\(^2\), column moment of inertia of \(1.6 \times 10^{-5}\) m\(^4\), and girder moment of inertia of \(5.4 \times 10^{-5}\) m\(^4\). All supports are fixed. What are the resulting deformations and internal forces (axial force, shear force, and bending moment) within the frame?
        & \begin{minipage}[t][4.5cm][t]{0.32\textwidth}
            \centering
            \includegraphics[scale=0.36]{images/schematic_13.png}
        \end{minipage} \\
        \hline
        \textbf{14. }How does a two-story two-bay 2D frame, where the first bay has one story and the second bay has two stories, consisting of 5 vertical columns (\(4\) meters in height each) and 3 horizontal girders (\(6\) meters in length each), behave under the uniform vertical distributed load of \(1 \times 10^4\) N/m along each girder? Consider elastic material properties with Young's modulus of \(2 \times 10^{11}\) Pa, column cross-sectional area of \(2 \times 10^{-3}\) m\(^2\), girder cross-sectional area of \(6 \times 10^{-3}\) m\(^2\), column moment of inertia of \(1.6 \times 10^{-5}\) m\(^4\), and girder moment of inertia of \(5.4 \times 10^{-5}\) m\(^4\). All supports are fixed. What are the resulting deformations and internal forces (axial force, shear force, and bending moment) within the frame?
        & \begin{minipage}[t][4.5cm][t]{0.32\textwidth}
            \centering
            \includegraphics[scale=0.36]{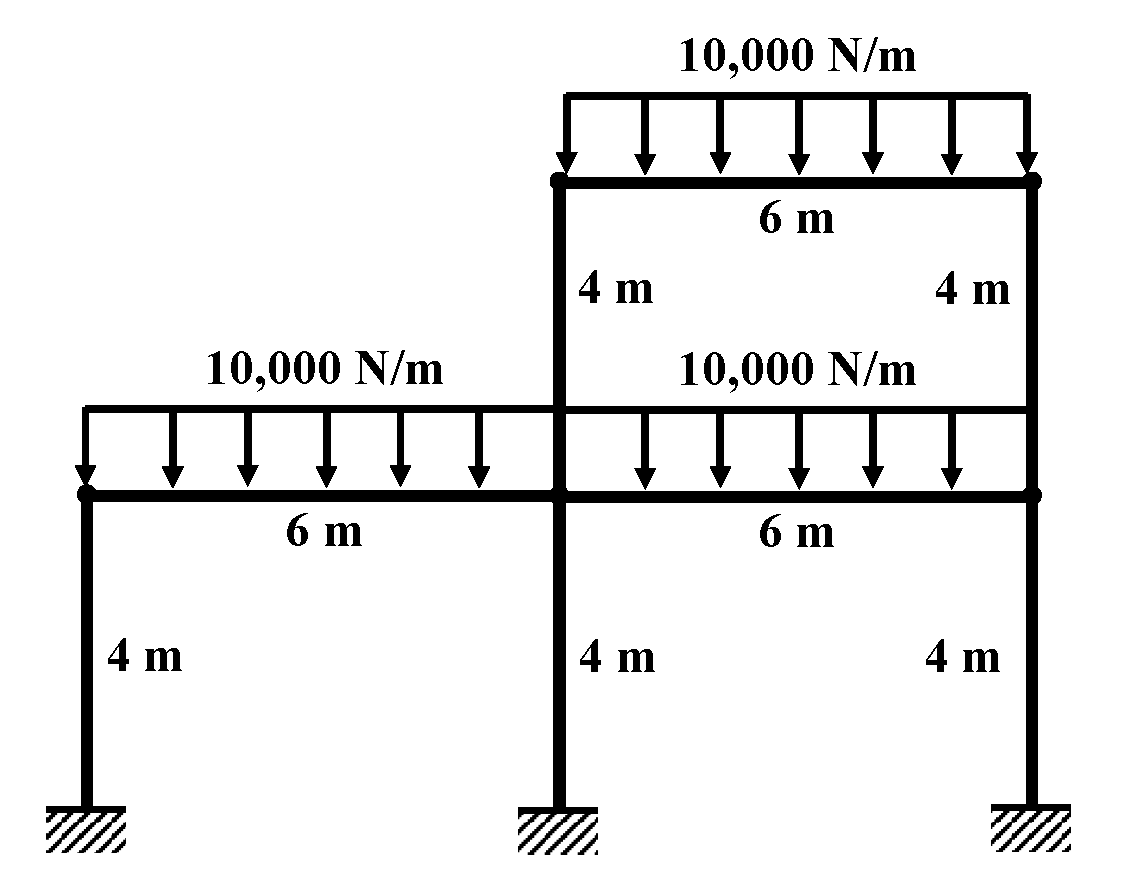}
        \end{minipage} \\
        \hline
        \textbf{15. }How does a simple 2D frame, consisting of two vertical columns (\(4 \times 10^0\) meters in height) with a spacing of (\(8 \times 10^0\)) meters between the two columns, have two identical diagonal members forming the roof in the middle of the columns, where the vertical height from the top of the columns to the peak of the roof is (\(3 \times 10^0\)) meters, behave under the effect of a horizontal point load of (\(2 \times 10^3\)) N at the left column? Consider elastic material properties with a Young's modulus \(E\) of (\(2 \times 10^{11}\)) Pa, column cross-sectional area of (\(2 \times 10^{-3}\)) m\(^2\), diagonal member cross-sectional area of (\(6 \times 10^{-3}\)) m\(^2\), column moment of inertia of (\(1.6 \times 10^{-5}\)) m\(^4\), and diagonal member moment of inertia of (\(5.4 \times 10^{-5}\)) m\(^4\). All supports are fixed. What are the resulting deformations and internal forces (axial, shear, and moment) in the frame?
        & \begin{minipage}[t][4.5cm][t]{0.32\textwidth}
            \centering
            \includegraphics[scale=0.4]{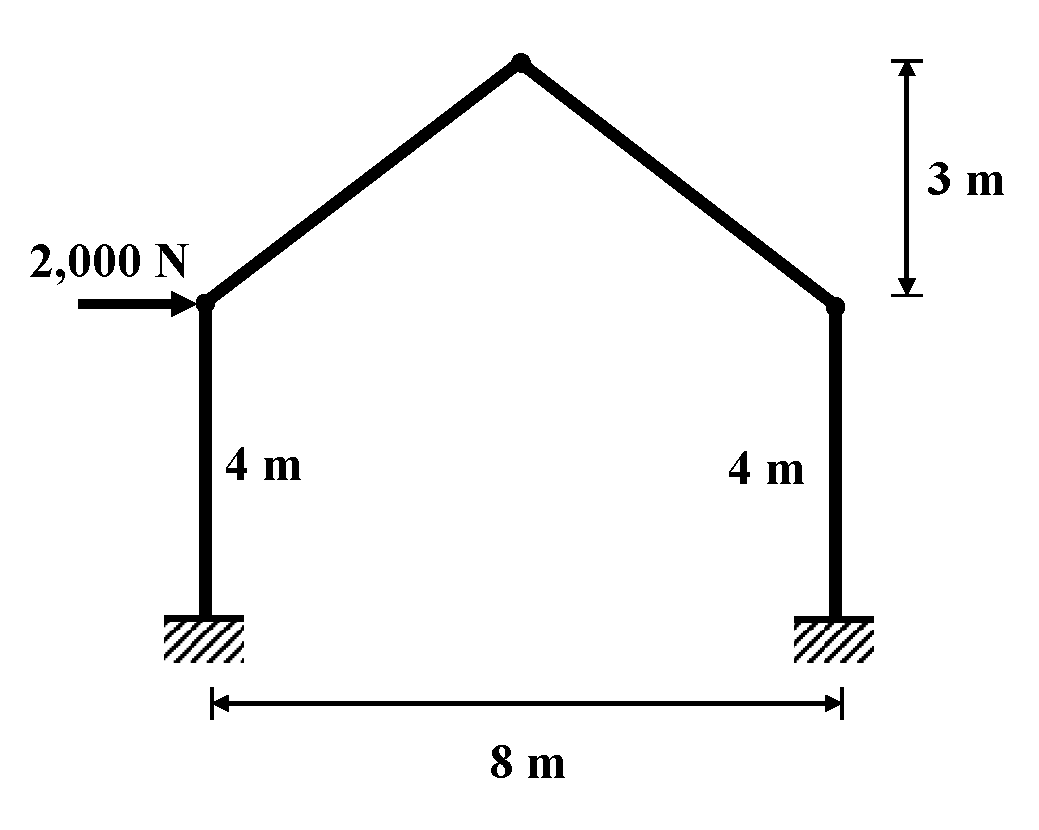}
        \end{minipage} \\
        \hline
        \textbf{16. }How does a simple 2D frame, consisting of two vertical columns (\(4 \times 10^0\) meters in height) with a spacing of (\(8 \times 10^0\)) meters between the two columns, have two identical diagonal members forming the roof in the middle of the columns, where the vertical height from the top of the columns to the peak of the roof is (\(3 \times 10^0\)) meters, behave under the uniform distributed load of (\(1 \times 10^4\)) N/m along each diagonal member? The direction of the distributed load is inward. Considering elastic material properties with a Young's modulus \(E\) of (\(2 \times 10^{11}\)) Pa, column cross-sectional area of (\(2 \times 10^{-3}\)) m\(^2\), diagonal member cross-sectional area of (\(6 \times 10^{-3}\)) m\(^2\), column moment of inertia of (\(1.6 \times 10^{-5}\)) m\(^4\), and diagonal member moment of inertia of (\(5.4 \times 10^{-5}\)) m\(^4\). All supports are fixed. What are the resulting deformations and internal forces (axial, shear, and moment) in the frame?
        & \begin{minipage}[t][4.5cm][t]{0.32\textwidth}
            \centering
            \includegraphics[scale=0.4]{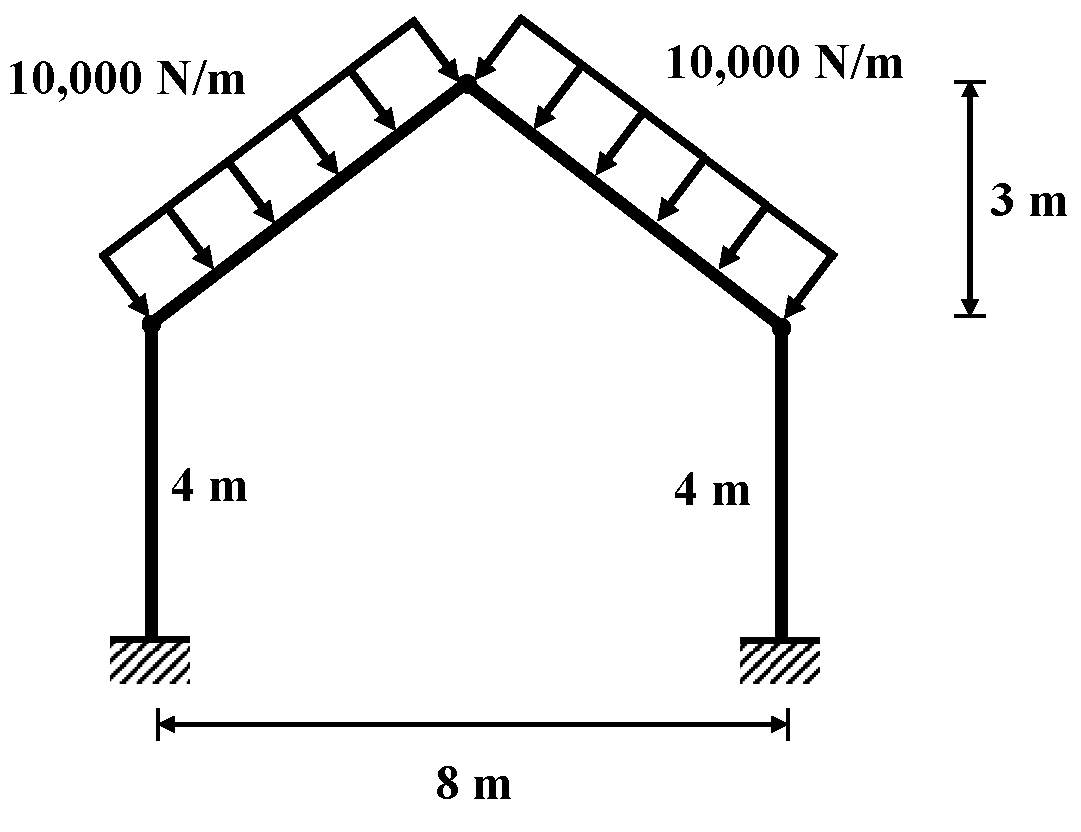}
        \end{minipage} \\
        \hline
    \end{tabularx}
\end{table}

% Fifth part of the table
\begin{table}[h]
    \centering
    \renewcommand{\arraystretch}{1.5}
    \caption{Additional Problem Descriptions and Ground Truth Schematics}
    \label{schematic_5}
    \setlength{\extrarowheight}{2pt}
    \renewcommand{\tabularxcolumn}[1]{>{\hsize=\dimexpr\hsize+2\tabcolsep\relax\justifying\arraybackslash}m{#1}}
    \begin{tabularx}{\textwidth}{|X|>{\centering\arraybackslash}m{0.32\textwidth}|} 
        \hline
        \textbf{Problem description} & \textbf{Ground truth} \\
        \hline
        \textbf{17. }How does a simple 2D frame, consisting of two vertical columns (\(4 \times 10^0\) meters in height), one horizontal girder (\(6 \times 10^0\) meters in length) and two diagonal members forming the braces (one node of the diagonal member is connected to the top of the column and another node is connected to the ground), one is on the left side of the left column and another is on the right side of the right column, where the horizontal length from the top of the column to the support of the diagonal member is (\(4 \times 10^0\)) meters, behave under the effect of a horizontal point load of (\(2 \times 10^3\)) N at the left column? Consider elastic material properties with a Young's modulus \(E\) of (\(2 \times 10^{11}\)) Pa, column cross-sectional area of (\(2 \times 10^{-3}\)) m\(^2\), diagonal member cross-sectional area of (\(6 \times 10^{-3}\)) m\(^2\), girder cross-sectional area of (\(6 \times 10^{-3}\)) m\(^2\), column moment of inertia of (\(1.6 \times 10^{-5}\)) m\(^4\), diagonal member moment of inertia of (\(5.4 \times 10^{-5}\)) m\(^4\), girder moment of inertia of (\(5.4 \times 10^{-5}\)) m\(^4\). All supports are fixed. What are the resulting deformations and internal forces (axial, shear, and moment) in the frame?
        & \begin{minipage}[t][4.5cm][t]{0.32\textwidth}
            \centering
            \includegraphics[scale=0.35]{images/schematic_17.png}
        \end{minipage} \\
        \hline
        \textbf{18. }How does a simple 2D frame, consisting of two vertical columns (\(4 \times 10^0\) meters in height), one horizontal girder (\(6 \times 10^0\) meters in length) and two diagonal members forming the braces (one node of the diagonal member is connected to the top of the column and another node is connected to the ground), one is on the left side of the left column and another is on the right side of the right column, where the horizontal length from the top of the column to the support of the diagonal member is (\(4 \times 10^0\)) meters, behave under the uniform distributed load of (\(1 \times 10^4\)) N/m along the girder? Consider elastic material properties with a Young's modulus \(E\) of (\(2 \times 10^{11}\)) Pa, column cross-sectional area of (\(2 \times 10^{-3}\)) m\(^2\), diagonal member cross-sectional area of (\(6 \times 10^{-3}\)) m\(^2\), girder cross-sectional area of (\(6 \times 10^{-3}\)) m\(^2\), column moment of inertia of (\(1.6 \times 10^{-5}\)) m\(^4\), diagonal member moment of inertia of (\(5.4 \times 10^{-5}\)) m\(^4\), girder moment of inertia of (\(5.4 \times 10^{-5}\)) m\(^4\). All supports are fixed. What are the resulting deformations and internal forces (axial, shear, and moment) in the frame?
        & \begin{minipage}[t][4.5cm][t]{0.32\textwidth}
            \centering
            \includegraphics[scale=0.35]{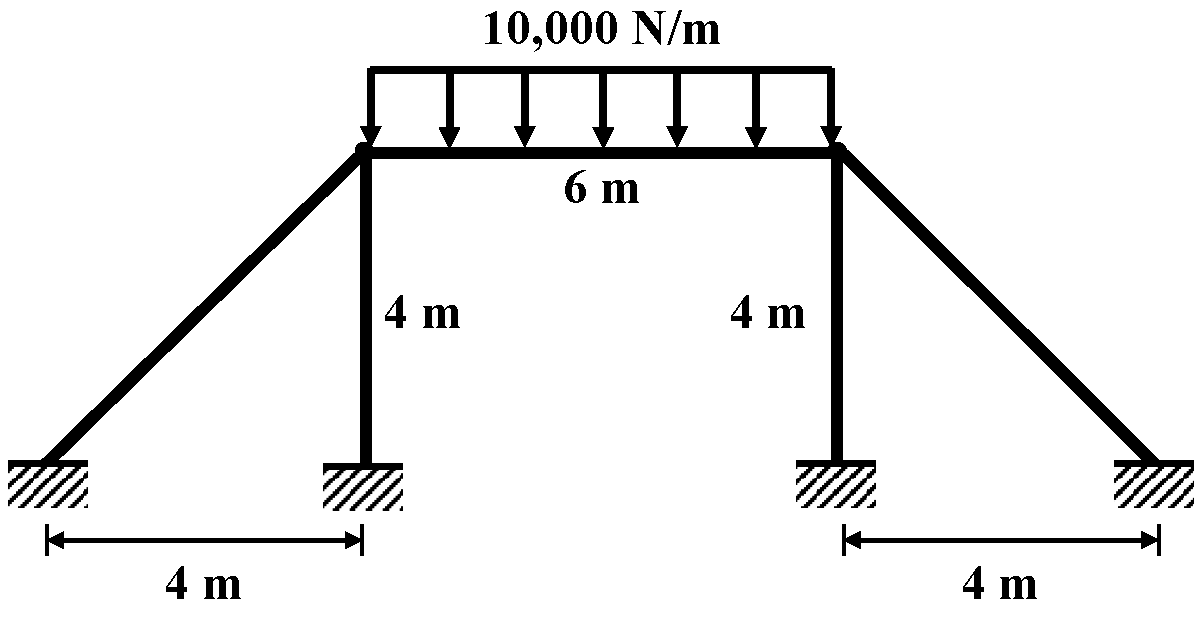}
        \end{minipage} \\
        \hline
        \textbf{19. }How does a simple 2D frame, consisting of two vertical columns (\(4 \times 10^0\) meters in height), one horizontal girder (\(6 \times 10^0\) meters in length) and two cantilever beams (\(2 \times 10^0\) meters in length) on both sides which are connected to the top of two columns, behave under the combined effects of two vertical point loads of (\(2 \times 10^3\)) N at the end of each cantilever beam on both sides? Consider elastic material properties with a Young's modulus \(E\) of (\(2 \times 10^{11}\)) Pa, column cross-sectional area of (\(2 \times 10^{-3}\)) m\(^2\), girder and cantilever beam cross-sectional area of (\(6 \times 10^{-3}\)) m\(^2\), column moment of inertia of (\(1.6 \times 10^{-5}\)) m\(^4\), and girder and cantilever beam moment of inertia of (\(5.4 \times 10^{-5}\)) m\(^4\). All supports are fixed. What are the resulting deformations and internal forces (axial, shear, and moment) in the frame?
        & \begin{minipage}[t][4.5cm][t]{0.32\textwidth}
            \centering
            \includegraphics[scale=0.49]{images/schematic_19.png}
        \end{minipage} \\
        \hline
        \textbf{20. }How does a simple 2D frame, consisting of two vertical columns (\(4 \times 10^0\) meters in height), one horizontal girder (\(6 \times 10^0\) meters in length) and two cantilever beams (\(2 \times 10^0\) meters in length) on both sides which are connected to the top of two columns, behave under the uniform vertical distributed load of (\(1 \times 10^4\)) N/m along the girder and two cantilever beams? Consider elastic material properties with a Young's modulus \(E\) of (\(2 \times 10^{11}\)) Pa, column cross-sectional area of (\(2 \times 10^{-3}\)) m\(^2\), girder and cantilever beam cross-sectional area of (\(6 \times 10^{-3}\)) m\(^2\), column moment of inertia of (\(1.6 \times 10^{-5}\)) m\(^4\), and girder and cantilever beam moment of inertia of (\(5.4 \times 10^{-5}\)) m\(^4\). All supports are fixed. What are the resulting deformations and internal forces (axial, shear, and moment) in the frame?
        & \begin{minipage}[t][4.5cm][t]{0.32\textwidth}
            \centering
            \includegraphics[scale=0.51]{images/schematic_20.png}
        \end{minipage} \\
        \hline
    \end{tabularx}
\end{table}

\clearpage
\renewcommand{\thesection}{Appendix \Alph{section}}  
\section{ICL Template} \label{icltemplate}
\begin{figure}[h]
    \centering
    \includegraphics[width=\textwidth, trim=5 80 5 0, clip]{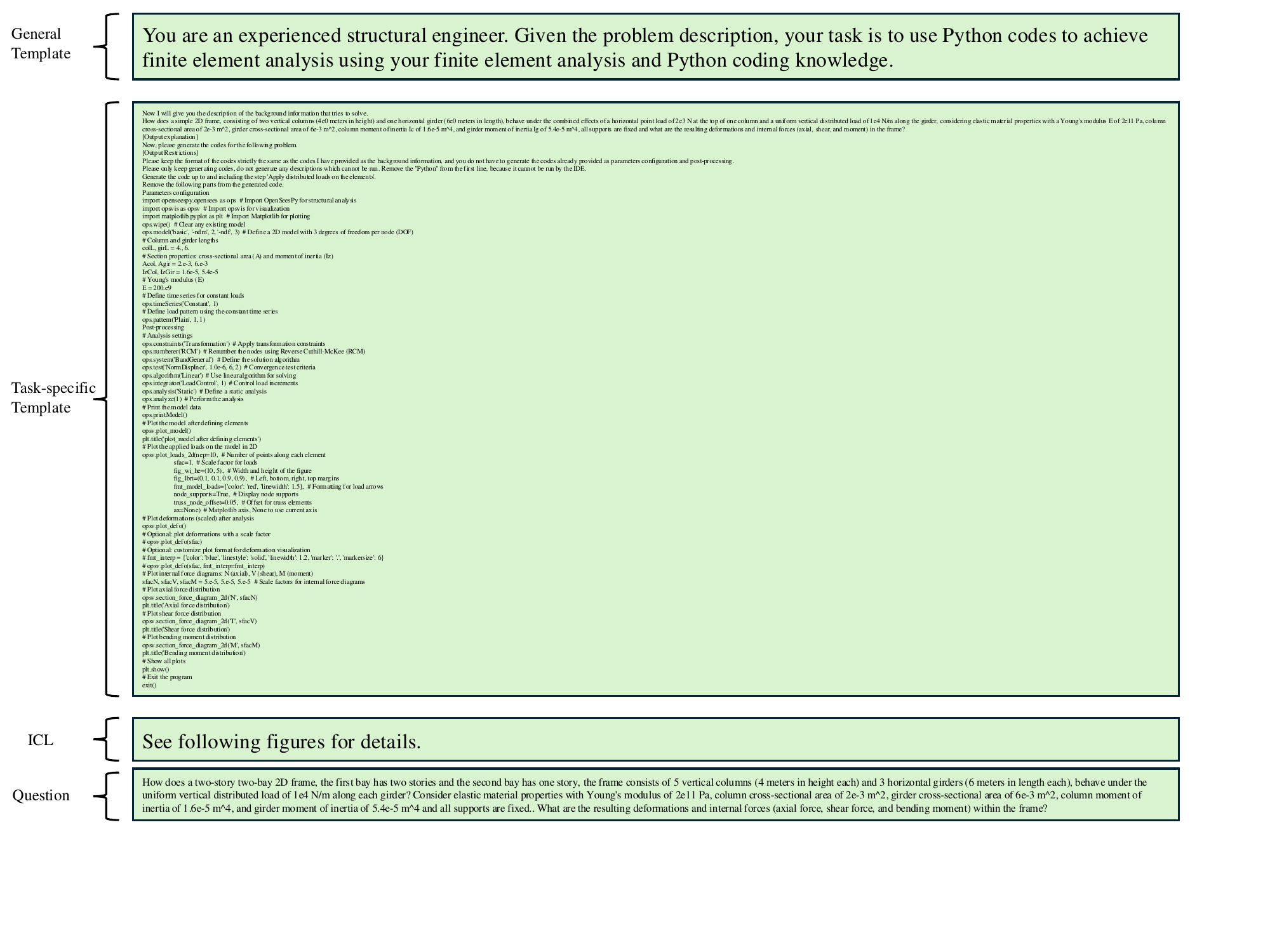}
    \caption{An ICL prompt example for solving example 12}
    \label{fig:icl_1}
\end{figure}

\clearpage

\begin{figure}[h]
    \centering
    \includegraphics[width=\textwidth, trim=5 0 5 0, clip]{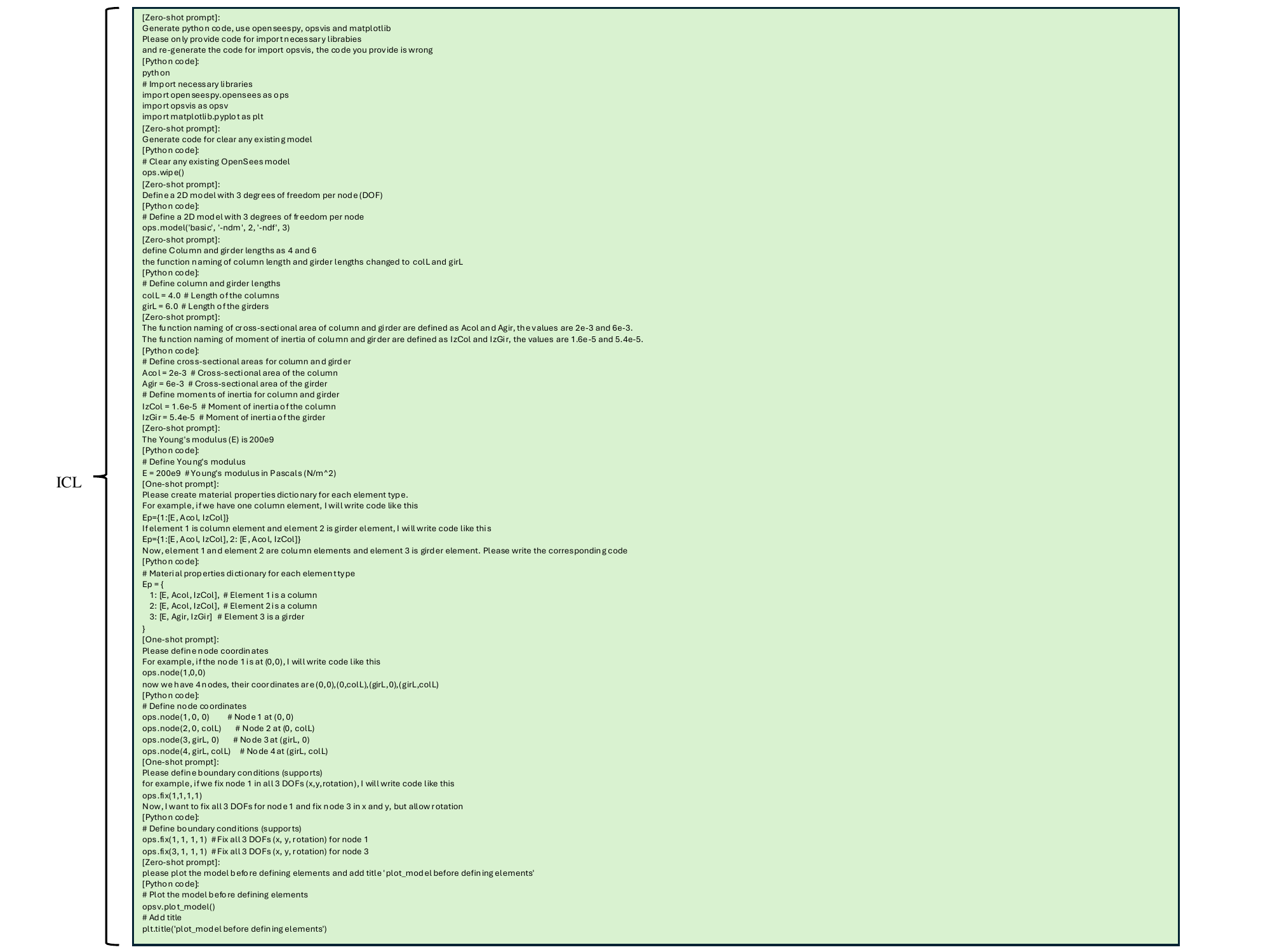}
    \caption{ICL detail 1}
    \label{fig:icl_2}
\end{figure}

\clearpage

\begin{figure}[h]
    \centering
    \includegraphics[width=\textwidth, trim=5 0 5 0, clip]{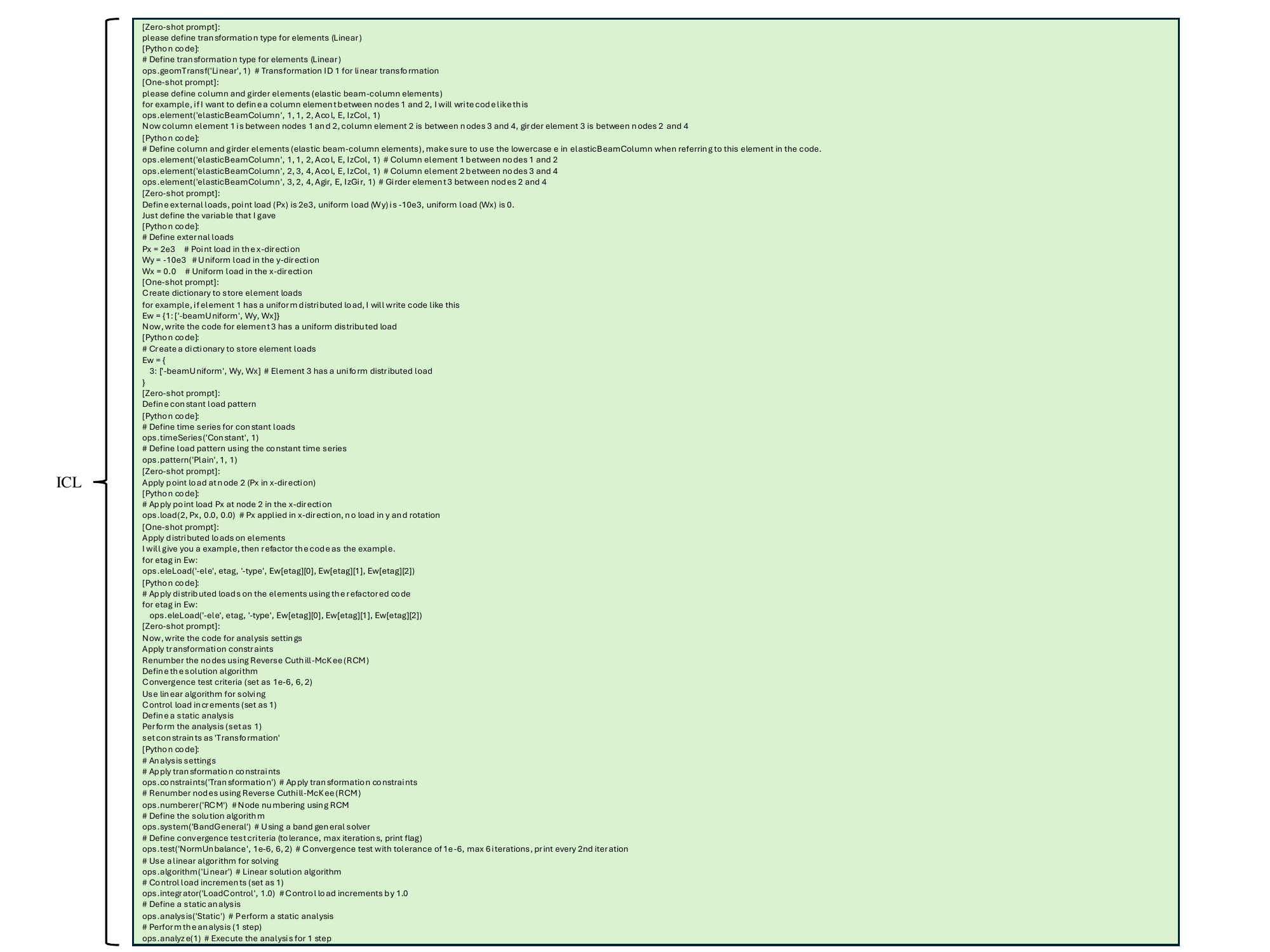}
    \caption{ICL detail 2}
    \label{fig:icl_3}
\end{figure}

\clearpage

\begin{figure}[h]
    \centering
    \includegraphics[width=\textwidth, trim=5 50 5 0, clip]{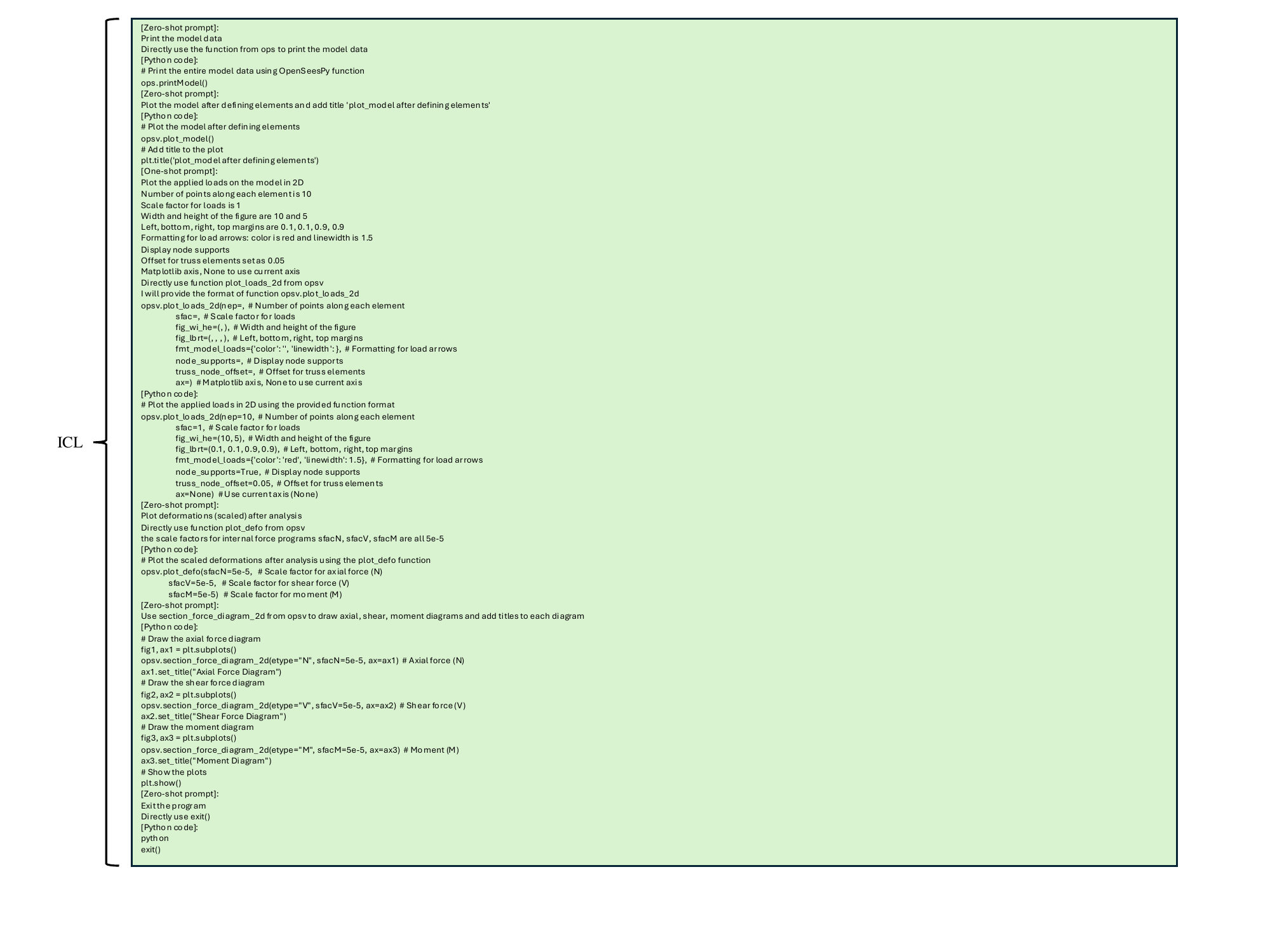}
    \caption{ICL detail 3}
    \label{fig:icl_4}

    \centering
    \includegraphics[width=\textwidth, trim=5 150 5 0, clip]{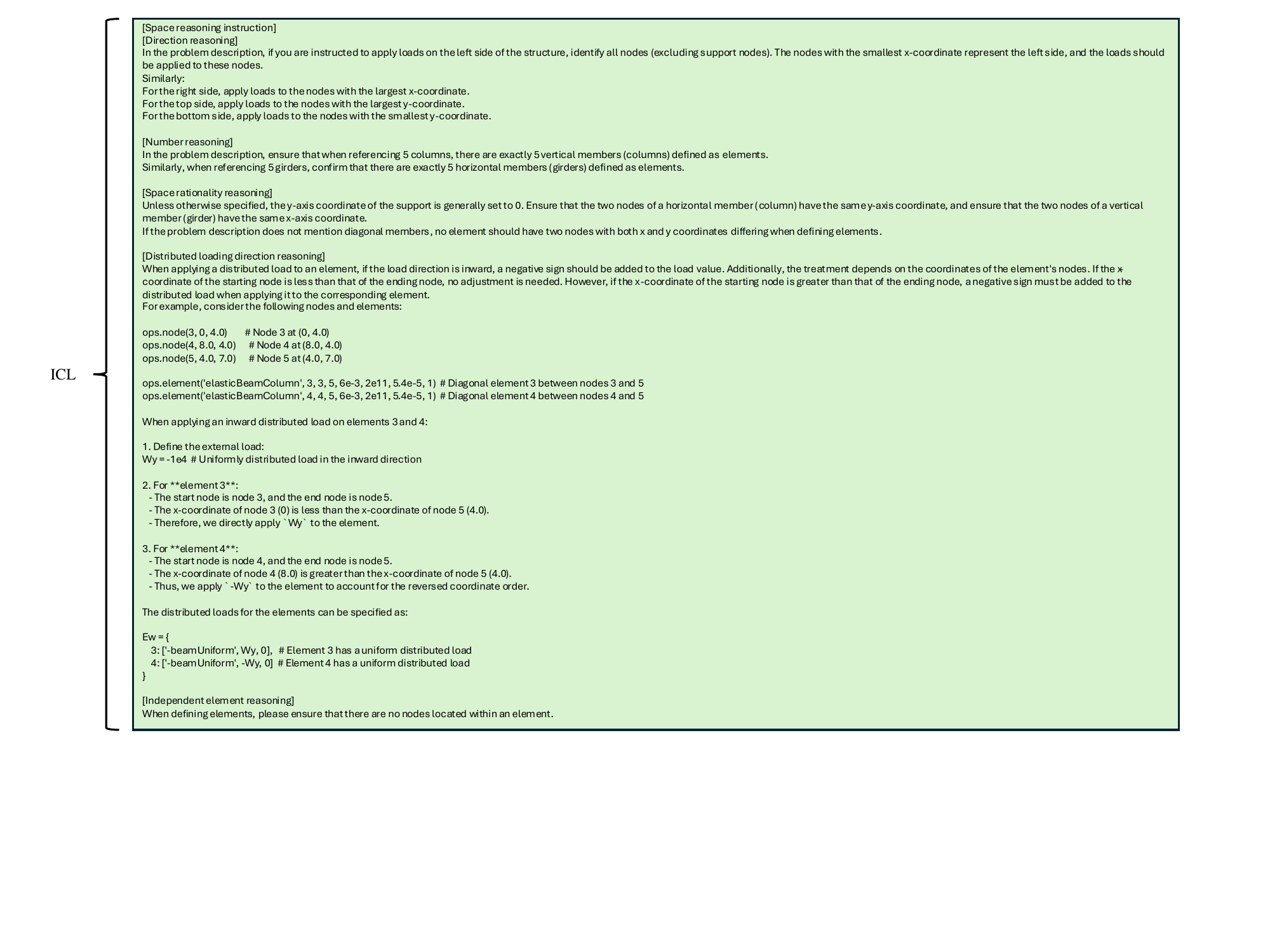}
    \caption{ICL detail 4}
    \label{fig:icl_5}
\end{figure}

\clearpage
\renewcommand{\thesection}{Appendix \Alph{section}}  
\section{Baseline} \label{baseline}

\begin{table}[ht]
\centering
\caption{\textbf{Failure case 1:GPT-4o's response without providing ICL template}}
\end{table}

\noindent\rule{\linewidth}{0.4pt}

\textbf{Question}: Solve the following problem: How does a simple 2D frame, consisting of two vertical columns (4e0 meters in height) and one horizontal girder (6e0 meters in length), behave under the combined effects of a horizontal point load of 2e3 N at the top of one column and a uniform vertical distributed load of 1e4 N/m along the girder, considering elastic material properties with a Young's modulus E of 2e11 Pa, column cross-sectional area of 2e-3 m\textasciicircum 2, girder cross-sectional area of 6e-3 m\textasciicircum 2, column moment of inertia Ic of 1.6e-5 m\textasciicircum 4, and girder moment of inertia Ig of 5.4e-5 m\textasciicircum 4, all supports are fixed and what are the resulting deformations and internal forces (axial, shear, and moment) in the frame?

\textbf{Answer}:
\section*{Problem Statement}

How does a simple 2D frame, consisting of two vertical columns (4.0 meters in height) and one horizontal girder (6.0 meters in length), behave under the combined effects of a horizontal point load of \(2 \times 10^3\,\mathrm{N}\) at the top of one column and a uniform vertical distributed load of \(1 \times 10^4\,\mathrm{N/m}\) along the girder? 

Assume the frame is made of an elastic material with the following properties:

\begin{itemize}
  \item Young's modulus: \(E = 2 \times 10^{11}\,\mathrm{Pa}\)
  \item Column cross-sectional area: \(A_c = 2 \times 10^{-3}\,\mathrm{m}^2\)
  \item Girder cross-sectional area: \(A_g = 6 \times 10^{-3}\,\mathrm{m}^2\)
  \item Column moment of inertia: \(I_c = 1.6 \times 10^{-5}\,\mathrm{m}^4\)
  \item Girder moment of inertia: \(I_g = 5.4 \times 10^{-5}\,\mathrm{m}^4\)
  \item All supports are fixed.
\end{itemize}

Determine the resulting deformations and internal forces (axial, shear, and moment) in the frame.

\section*{Qualitative Behavior}

\subsection*{Effect of Horizontal Load}

\begin{itemize}
  \item The frame experiences lateral sway due to the horizontal load.
  \item The top of the loaded column deflects horizontally.
  \item Axial forces develop in the girder and columns.
\end{itemize}

\subsection*{Effect of Vertical Uniform Load}

\begin{itemize}
  \item The girder bends downward, with maximum bending at midspan.
  \item Shear and moment reactions are induced at the column bases.
  \item Axial forces arise in the columns due to vertical load.
\end{itemize}

\section*{Deflection Estimates}

\subsection*{Horizontal Deflection at Top of Left Column}

Approximated using cantilever beam deflection formula:

\[
\delta_H = \frac{P H^3}{3 E I_c}
\]

Substituting the given values:

\[
\delta_H = \frac{2000 \times 4^3}{3 \times 2 \times 10^{11} \times 1.6 \times 10^{-5}} 
= \frac{128000}{9.6 \times 10^6} \approx 0.0133\,\mathrm{m} = 13.3\,\mathrm{mm}
\]

\subsection*{Vertical Deflection at Midspan of Girder}

Approximated using uniform load beam deflection formula:

\[
\delta_V = \frac{5 w L^4}{384 E I_g}
\]

\[
\delta_V = \frac{5 \times 10^4 \times 6^4}{384 \times 2 \times 10^{11} \times 5.4 \times 10^{-5}} 
= \frac{6.48 \times 10^7}{4.1472 \times 10^9} \approx 0.0156\,\mathrm{m} = 15.6\,\mathrm{mm}
\]

\section*{Internal Force Estimates}

\subsection*{Girder}

\begin{itemize}
  \item Maximum bending moment:
  \[
  M_{\text{max}} = \frac{w L^2}{8} = \frac{10^4 \times 6^2}{8} = 45{,}000\,\mathrm{Nm}
  \]
  \item Shear force at supports:
  \[
  V = \frac{w L}{2} = 30{,}000\,\mathrm{N}
  \]
  \item Axial force due to sway: non-zero
\end{itemize}

\subsection*{Columns}

\begin{itemize}
  \item Bending moments at top and bottom due to lateral sway and fixed supports
  \item Axial forces: one column in compression, the other in tension
  \item Shear forces result from both horizontal and vertical loading
\end{itemize}

\section*{Summary Table}

\begin{tabular}{@{}ll@{}}
\toprule
\textbf{Quantity} & \textbf{Approximate Value} \\ \midrule
Horizontal deflection (top of column) & \( \approx 13.3\,\mathrm{mm} \) \\
Vertical deflection (midspan of girder) & \( \approx 15.6\,\mathrm{mm} \) \\
Max bending moment in girder & \( \approx 45{,}000\,\mathrm{Nm} \) \\
Shear force at girder supports & \( \approx 30{,}000\,\mathrm{N} \) \\
Axial force in girder & Non-zero due to sway \\
Column moments & Significant at top and base \\
Axial forces in columns & Opposite (one in tension, one in compression) \\
\bottomrule
\end{tabular}

\section*{Conclusion}

The frame experiences both vertical and horizontal displacements due to combined loads. Bending, shear, and axial forces are generated in both the girder and columns. For exact analysis, matrix stiffness method or finite element software (e.g., SAP2000, ANSYS) should be used.

\noindent\rule{\linewidth}{0.4pt}

\clearpage
\begin{table}[ht]
\centering
\caption{\textbf{Failure case 2: GPT-4o's response without providing ICL template}}
\end{table}

\noindent\rule{\linewidth}{0.4pt}

\textbf{Question}: Solve the following problem: How does a simple 2D frame, consisting of two vertical columns (4e0 meters in height) and one horizontal girder (6e0 meters in length), behave under the combined effects of a horizontal point load of 2e3 N at the top of one column and a uniform vertical distributed load of 1e4 N/m along the girder, considering elastic material properties with a Young's modulus E of 2e11 Pa, column cross-sectional area of 2e-3 m\textasciicircum 2, girder cross-sectional area of 6e-3 m\textasciicircum 2, column moment of inertia Ic of 1.6e-5 m\textasciicircum 4, and girder moment of inertia Ig of 5.4e-5 m\textasciicircum 4, all supports are fixed and what are the resulting deformations and internal forces (axial, shear, and moment) in the frame? \textbf{Use code}.

\textbf{Answer}:

\section*{Problem Description}

We analyze a simple 2D frame consisting of:
\begin{itemize}
  \item Two vertical columns, each of height $4\,\text{m}$.
  \item One horizontal girder of length $6\,\text{m}$.
  \item A horizontal point load of $2 \times 10^3\,\text{N}$ at the top of the left column.
  \item A uniform vertical distributed load of $1 \times 10^4\,\text{N/m}$ applied on the girder.
\end{itemize}

Material and section properties:
\begin{itemize}
  \item Young's Modulus: $E = 2 \times 10^{11}\,\text{Pa}$
  \item Cross-sectional area of columns: $2 \times 10^{-3}\,\text{m}^2$
  \item Cross-sectional area of girder: $6 \times 10^{-3}\,\text{m}^2$
  \item Moment of inertia of columns: $1.6 \times 10^{-5}\,\text{m}^4$
  \item Moment of inertia of girder: $5.4 \times 10^{-5}\,\text{m}^4$
\end{itemize}

All supports are fixed.

\section*{Python Code for Structural Analysis}

\begin{lstlisting}[language=Python]
import numpy as np
import matplotlib.pyplot as plt

# Frame geometry
H = 4e0
L = 6e0

# Loads
P_horizontal = 2e3
w_vertical = 1e4

# Material properties
E = 2e11

# Section properties
A_column = 2e-3
A_girder = 6e-3
I_column = 1.6e-5
I_girder = 5.4e-5

# Nodes: 0 (left base), 1 (left top), 2 (right top), 3 (right base)
ndof = 3 * 4
elements = [
    {'nodes': (0, 1), 'A': A_column, 'I': I_column, 'E': E},
    {'nodes': (1, 2), 'A': A_girder, 'I': I_girder, 'E': E},
    {'nodes': (2, 3), 'A': A_column, 'I': I_column, 'E': E}
]

coords = {
    0: [0, 0],
    1: [0, H],
    2: [L, H],
    3: [L, 0]
}

K_global = np.zeros((ndof, ndof))
F_global = np.zeros(ndof)

def element_stiffness(E, A, I, L, angle):
    c = np.cos(angle)
    s = np.sin(angle)
    k = np.array([
        [A*c**2, A*c*s, 0, -A*c**2, -A*c*s, 0],
        [A*c*s, A*s**2, 0, -A*c*s, -A*s**2, 0],
        [0, 0, 12*I/L**3, 0, 0, -6*I/L**2],
        [-A*c**2, -A*c*s, 0, A*c**2, A*c*s, 0],
        [-A*c*s, -A*s**2, 0, A*c*s, A*s**2, 0],
        [0, 0, -6*I/L**2, 0, 0, 4*I/L]
    ])
    return E/L * k

for element in elements:
    n1, n2 = element['nodes']
    x1, y1 = coords[n1]
    x2, y2 = coords[n2]
    dx = x2 - x1
    dy = y2 - y1
    length = np.sqrt(dx**2 + dy**2)
    angle = np.arctan2(dy, dx)

    k_local = element_stiffness(element['E'], element['A'], element['I'], length, angle)
    dof_map = [3*n1, 3*n1+1, 3*n1+2, 3*n2, 3*n2+1, 3*n2+2]

    for i in range(6):
        for j in range(6):
            K_global[dof_map[i], dof_map[j]] += k_local[i, j]

F_global[4] -= w_vertical * L / 2
F_global[7] -= w_vertical * L / 2
F_global[3] += P_horizontal

fixed_dofs = [0, 1, 2, 9, 10, 11]
free_dofs = list(set(range(ndof)) - set(fixed_dofs))

K_ff = K_global[np.ix_(free_dofs, free_dofs)]
F_f = F_global[free_dofs]
u_f = np.linalg.solve(K_ff, F_f)

u = np.zeros(ndof)
u[free_dofs] = u_f
\end{lstlisting}

\section*{Nodal Displacement Results}

\begin{tabular}{cccc}
\toprule
\textbf{Node} & \textbf{u (m)} & \textbf{v (m)} & \textbf{$\theta$ (rad)} \\
\midrule
0 & $0.000000 \times 10^{0}$ & $0.000000 \times 10^{0}$ & $0.0$ \\
1 & $-5.334188 \times 10^{27}$ & $3.266248 \times 10^{11}$ & $0.0$ \\
2 & $-5.334188 \times 10^{27}$ & $-3.266248 \times 10^{11}$ & $0.0$ \\
3 & $0.000000 \times 10^{0}$ & $0.000000 \times 10^{0}$ & $0.0$ \\
\bottomrule
\end{tabular}

\noindent\rule{\linewidth}{0.4pt}

\begin{table}[htbp]
\centering
\caption{Ground Truth Solution of Example 1}
\label{tab:gt_solution}
\renewcommand{\arraystretch}{1.2}
\begin{tabularx}{\textwidth}{>{\bfseries}l X}
\toprule
\multicolumn{2}{l}{\textbf{Node Displacements}} \\
Node 1 & (0, 0, 0) \\
Node 2 & (0.00203106, -0.000293798, -0.00458888) \\
Node 3 & (0, 0, 0) \\
Node 4 & (0.00199962, -0.000306202, 0.0042402) \\
\addlinespace
\multicolumn{2}{l}{\textbf{Element End Forces (P, V, M)}} \\
Element 1 - End 1 & (29379.8, -4288.01, -4904.93) \\
Element 1 - End 2 & (-29379.8, 4288.01, -12247.1) \\
Element 2 - End 1 & (30620.2, 6288.01, 9183.87) \\
Element 2 - End 2 & (-30620.2, -6288.01, 15968.2) \\
Element 3 - End 1 & (6288.01, 29379.8, 12247.1) \\
Element 3 - End 2 & (-6288.01, 30620.2, -15968.2) \\
\bottomrule
\end{tabularx}
\end{table}

\begin{figure}[htbp]
    \centering
    \includegraphics[width=0.75\textwidth, trim=0 250 0 250, clip]{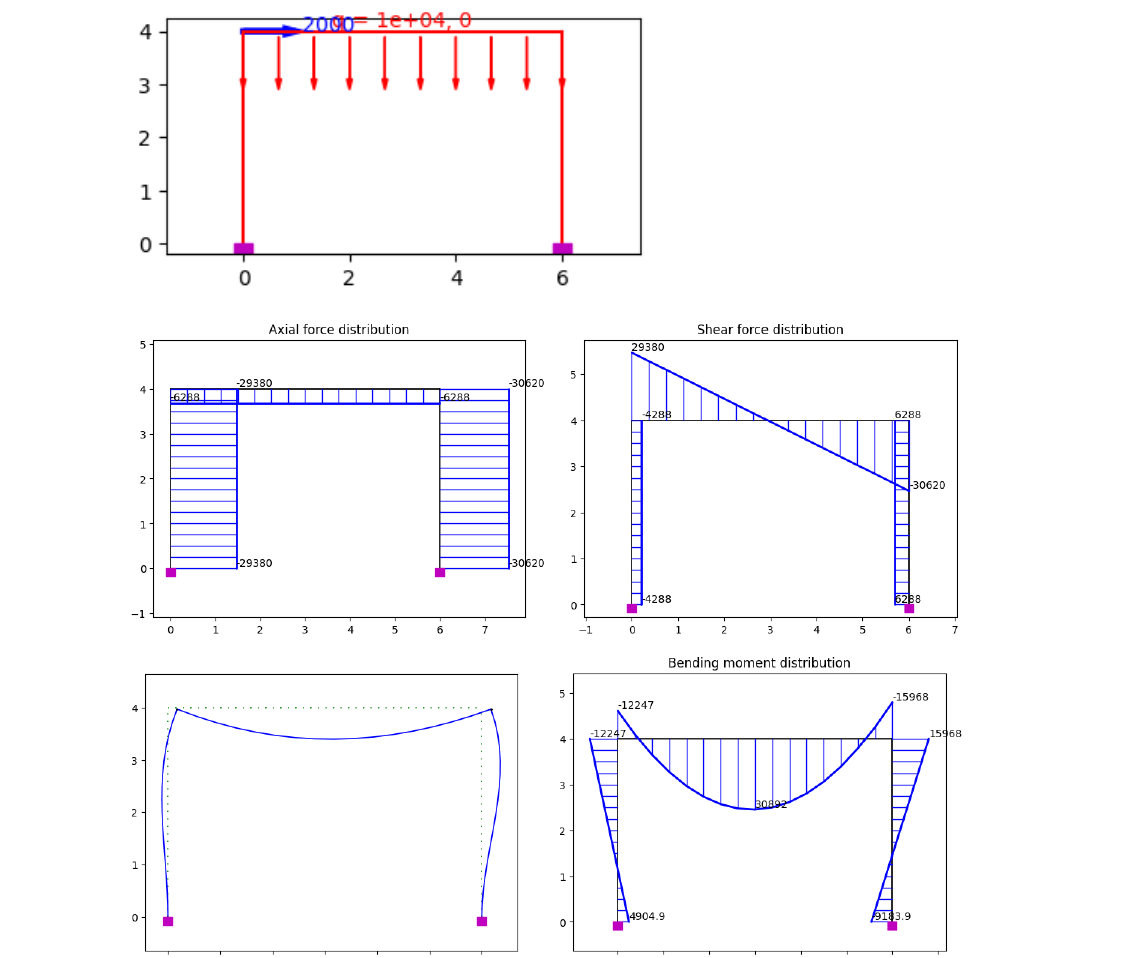}
    \caption{Ground Truth Solution of Example 1}
    \label{fig:groundtruth}
\end{figure}

\end{document}